\documentclass[pra,twocolumn,showpacs,preprintnumbers,amsmath,amssymb,floatfix]{revtex4}
\usepackage{graphicx}
\usepackage{dcolumn}
\usepackage{bm}

\begin{document}

\title{Momentum distribution and ordering in mixtures of ultracold light 
and heavy fermionic atoms}

\author{M. M. Ma\'ska}
 \affiliation{Department of Theoretical Physics, Institute of Physics, University of Silesia, 
PL-40007 Katowice, Poland}
\author{R. Lema\'nski}
\affiliation{Institute of Low Temperature and Structure Research, 
Polish Academy of Science, PL-50422 Wroc{\l}aw, Poland}
\author{C. J. Williams}
\affiliation{Joint Quantum Institute, National Institute of Standards and Technology, 
and University of Maryland, 100 Bureau Drive, Stop 8423, Gaithersburg, 
Maryland 20899-8423, USA}
\author{J. K. Freericks}
\affiliation{Department of Physics, Georgetown University, Washington, DC 20057, USA}

\date{\today}

\begin{abstract}
The momentum distribution is one of the most important quantities which provides
information about interactions in many-body systems. At the same time
it is a quantity that can easily be accessed in experiments on ultracold
atoms. In this paper, we consider mixtures of light and heavy fermionic 
atoms in an optical lattice described effectively by the Falicov-Kimball
model. Using a Monte Carlo method, we study how different ordered density-wave phases can be 
detected by measurement of the momentum distribution of the light atoms.
We also demonstrate that ordered phases can be seen in Bragg scattering 
experiments. Our results indicate that the main factor that determines the 
momentum distribution of the light atoms is the trap confinement. On the
other hand, the pattern formed by the heavy atoms seen in the Bragg 
scattering experiments is very sensitive to the temperature and possibly 
can be used in low-temperature thermometry.
\end{abstract}

\pacs{03.75.Ss, 67.85.-d, 67.85.Pq, 71.10.Fd}
\maketitle

\section{Introduction}
It is well known that the equilibrium momentum distribution of free noninteracting fermions
is described by the Fermi-Dirac distribution function. When interactions are turned on, the momentum distribution is changed due to these interactions. Similarly, if the noninteracting gas is
confined in a harmonic trap, the distribution can also be calculated explicitly \cite{rigol},
however in the limit of a large number of fermions it is convenient to use the semiclassical
Thomas--Fermi approximation \cite{butt} or other semiclassical approaches \cite{march}.
The situation becomes much more complicated
if the interaction between fermions cannot be neglected. Then, numerical methods
usually need to be applied, especially when the system is in a trap. On the other hand, the momentum distribution function
is a quantity that can be directly accessed in experiments with ultracold atomic
gases and therefore it is of a particular importance in trying to make contact between theory and experiment.

We examine the problem of a mixture of different mass fermionic atoms on an optical lattice at low temperature. Since most stable heavy isotopes of alkalies are bosonic, we envision mixtures of Li$^6$ with K$^{40}$, or light Li$^6$ or K$^{40}$ mixed with heavy fermionic isotopes of Sr or Yb (it turns out 
that if the heavy particle is a boson with strong enough intraspecies repulsion, and one is at a low enough temperature, then the statistics of the heavy particle does not enter into our model, as it appears effectively like a hardcore object).  We have already performed numerical calculations on this system and shown that one can see interesting phenomena similar to viscous fingering as one tunes the trap curvature for the light species and moves from a phase separated state with the heavies on the outside to one with the heavies on the inside~\cite{PRL}. Interesting ordered density-wave patterns appear at low temperature in these mixtures.  The question is, how do we detect the presence of such order with current experimentally available techniques?  In situations where {\it in situ} imaging with single-site precision is available~\cite{Greiner}, one would simply look for the ordered phases directly, just as they appear in the Monte Carlo snapshots of a particular configuration of the atoms.  But can one see effects of the density-wave ordering in a time-of-flight image, or via direct Bragg scattering of light off of the density-wave pattern?  We answer these questions here.

In Section II, we introduce the model and the techniques used to solve for the equilibrium properties of the mixtures of fermions in a harmonic trap.  In Section III, we present our numerical results.  Conclusions follow in Section IV.

\section{The model and method}
A mixture of light and heavy fermionic atoms in a harmonic trap (each in one and only one hyperfine atomic state and hence acting like a spinless fermion), under the assumption 
that the quantum-mechanical effects of the hopping of the heavy atoms can be neglected, is described by the 
Falicov--Kimball Hamiltonian \cite{FK,Ziegler}
\begin{eqnarray}
{\cal H}=&-&J\sum_{\langle i,j\rangle} \left(c^\dagger_ic^{}_j + c^\dagger_jc^{}_i\right) 
+\sum_i\left(V_i-\mu\right)c^\dagger_ic^{}_i \nonumber \\
&+& \sum_i\left(V^h_i-\mu^h\right)w_i + U\sum_ic^\dagger_ic^{}_jw_i,
\label{ham}
\end{eqnarray}
where $c^\dagger_i$  is an operator that creates a light fermionic atom
at site $i$ and $U$ is the on--site interspecies interaction potential. The symbol $V_i$ is the light particle trap potential, and $\mu$ is its chemical potential.  $V_i^h$ is the trap for the heavy atoms and $\mu^h$ is its respective chemical potential.  The symbol $w_i=0$ or 1 is the number operator of the heavy particles, which can be treated as a classical variable since the heavy particles do not hop. The same 
Hamiltonian can also describe Fermi-Bose mixtures, provided the bosonic 
atoms are the heavy (localized) ones and a strong repulsion between them prevents them from
occupying one site by two bosons (hard core bosons).
This model can easily be
extended to describe a system with soft-core bosons. In such a case $w_i=0,1,2,\ldots$
and an additional term describing the boson-boson on-site interaction
\begin{equation}
{\cal H}_{\rm B-B}=\frac{1}{2}U_{\rm B-B}\sum_iw_i(w_i-1)
\end{equation}
has to be added to the Hamiltonian in Eq.~(\ref{ham})~\cite{vollhardt,iskin}.
The trap potentials for the light ($V_i$) and heavy ($V^h_i$) atoms are given by
\begin{equation}
V_i=\frac{J}{R^2}\left(x_i^2+y_i^2\right),\ \ \ V^h_i=\frac{J}{\left(R^h\right)^2}\left(x_i^2+y_i^2\right)
\label{trap}
\end{equation}
where $(x_i,y_i)$ is the position of site $i$ 
 ($a$ is the lattice constant). The steepness of the potential confining the light atoms
varies from $R=30a$ to $R=12.9a$, whereas it is fixed for the
heavy atoms with $R^h=30a$. The chemical potentials $\mu$ and $\mu^h$ have been 
introduced in order to control the number of the light and heavy atoms, respectively.

The model is solved by means of a variation of the Monte Carlo (MC) method. The method is 
based on the classical Metropolis algorithm modified in such a way that systems
with both quantum and classical degrees of freedom can be simulated \cite{MC}.
In each MC step, a new configuration of the heavy atoms is generated. Then, 
the Hamiltonian in Eq.~(\ref{ham}) is numerically diagonalized to yield all the eigenenergies 
and eigenstates of the light atoms for the trial configuration of the heavy atoms. The new configuration of the heavy atoms is accepted according to the
same rules as in the Metropolis algorithm, but with the free energy of the light atoms
used for comparing energies instead of the internal energy. The results are then averaged over all the configurations 
generated during the entire MC run. Real-space configurations of the atoms for different 
model parameters (interaction, shape of the trapping potential) at different temperatures 
have been studied in Ref. \onlinecite{PRL}. In the present paper we use a similar 
method to investigate the momentum distribution and Bragg scattering spectra. Since diagonalization of the Hamiltonian in Eq.~(\ref{ham}) for a given configuration 
of the heavy atoms gives all the eigenstates of the light atoms, it allows one to also calculate 
the momentum distribution of the light atoms. Under the assumption that the optical lattice potential is deep enough that we can restrict to a single-band model, then
the field operator of these atoms $\Psi({\bm r})$ (expanded in terms of the Bloch wavefunctions for the lowest band) is given by
\begin{equation}
\Psi({\bm r})={\sum_{\bm k}}c_{\bm k}\Psi_{\bm k}({\bm r}),
\end{equation}
where $c_{\bm k}=1/N\sum_ic_i\exp\left(-i{\bm k}\cdot {\bm R}_i\right)$ and $\Psi_{\bm k}({\bm r})$
is the Bloch wavefunction. Expanding the Bloch wavefunction in terms of the Wannier wavefunctions of the lowest band $w({\bm r}-{\bm R}_i)$ which are localized about lattice site ${\bm R}_i$ yields
$\Psi_{\bm k}({\bm r})=1/\sqrt{N}\sum_i w({\bm r}-{\bm R}_i)\exp\left(i{\bm k}\cdot {\bm R}_i\right)$.
The summation runs over the first Brilliouin zone. Then, the free space momentum distribution can 
be calculated as \cite{wang}
\begin{equation}
n({\bm k})= |w({\bm k})|^2 c^\dagger_{\bm k} c^{}_{\bm k},
\end{equation} 
where $w({\bm k})$ is the Fourier transform of the Wannier function. 
Hence the momentum distribution can be approximated by
\begin{equation}
n({\bf k})=\frac{\left|w({\bm k})\right|^2}{N}
\sum_{i,j}\langle c^\dagger_i c^{}_j \rangle e^{i{\bm k}\cdot\left({\bm R}_i-{\bm R}_j\right)}
\label{eq7}
\end{equation}
in the single-band limit, with $N$ the number of lattice sites.

Here, the quantum mechanical expectation value $\langle\ldots\rangle$ is
calculated for a given configuration of the heavy atoms and therefore
$n({\bm k})$ has to be averaged over the configurations.

\begin{figure*}[htb]
\includegraphics[width=0.33\textwidth]{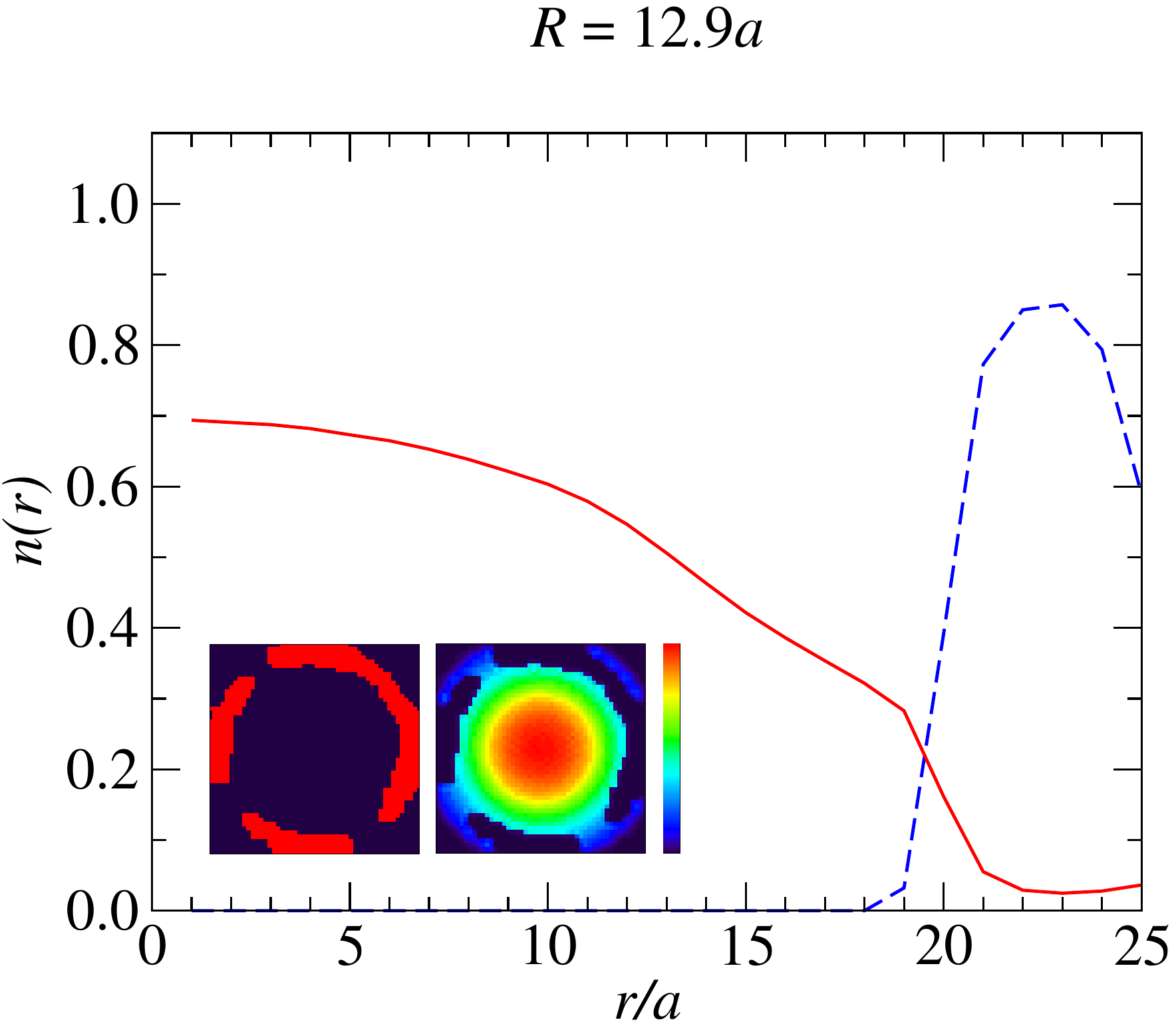}\includegraphics[width=0.33\textwidth]{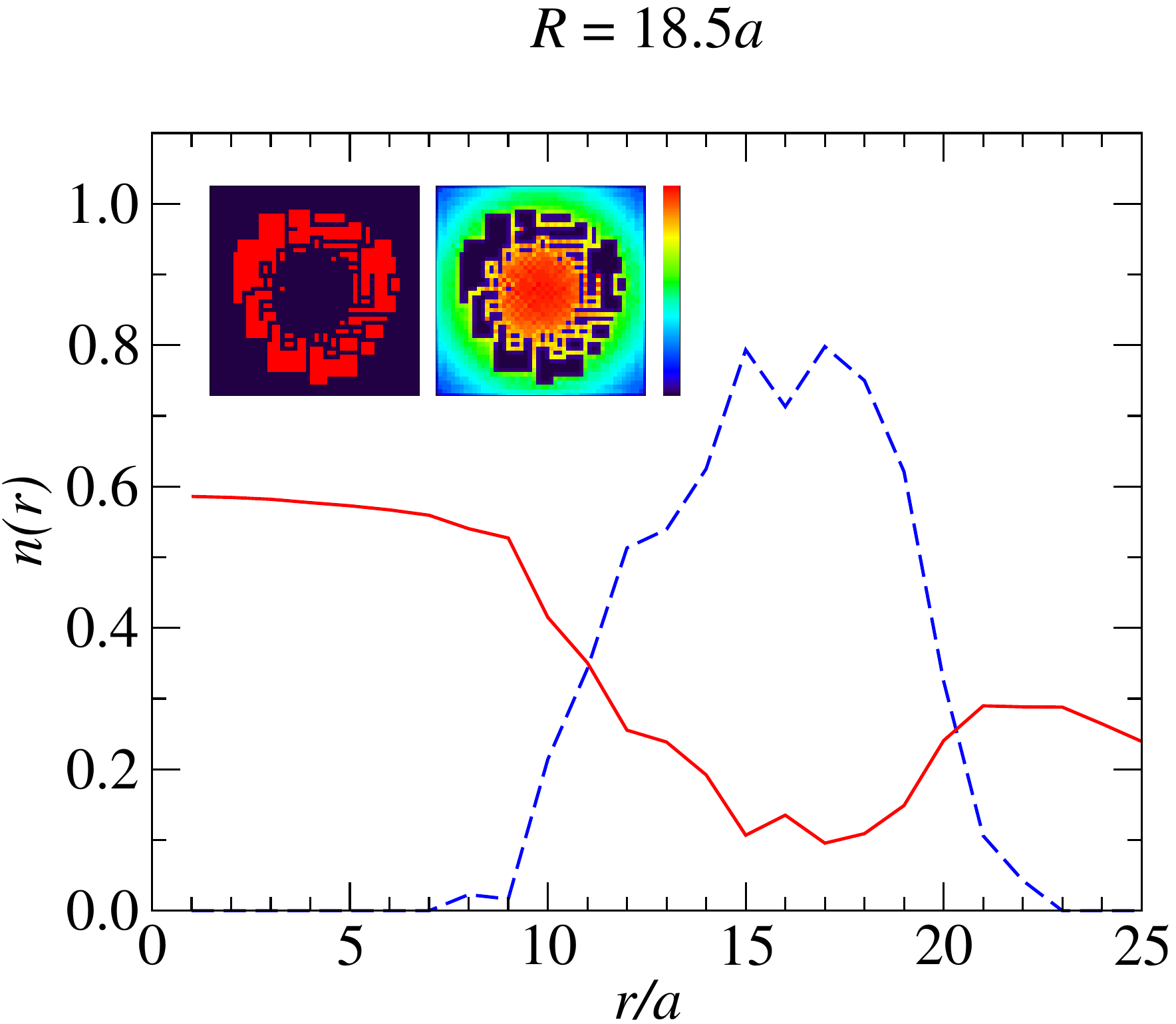}\includegraphics[width=0.33\textwidth]{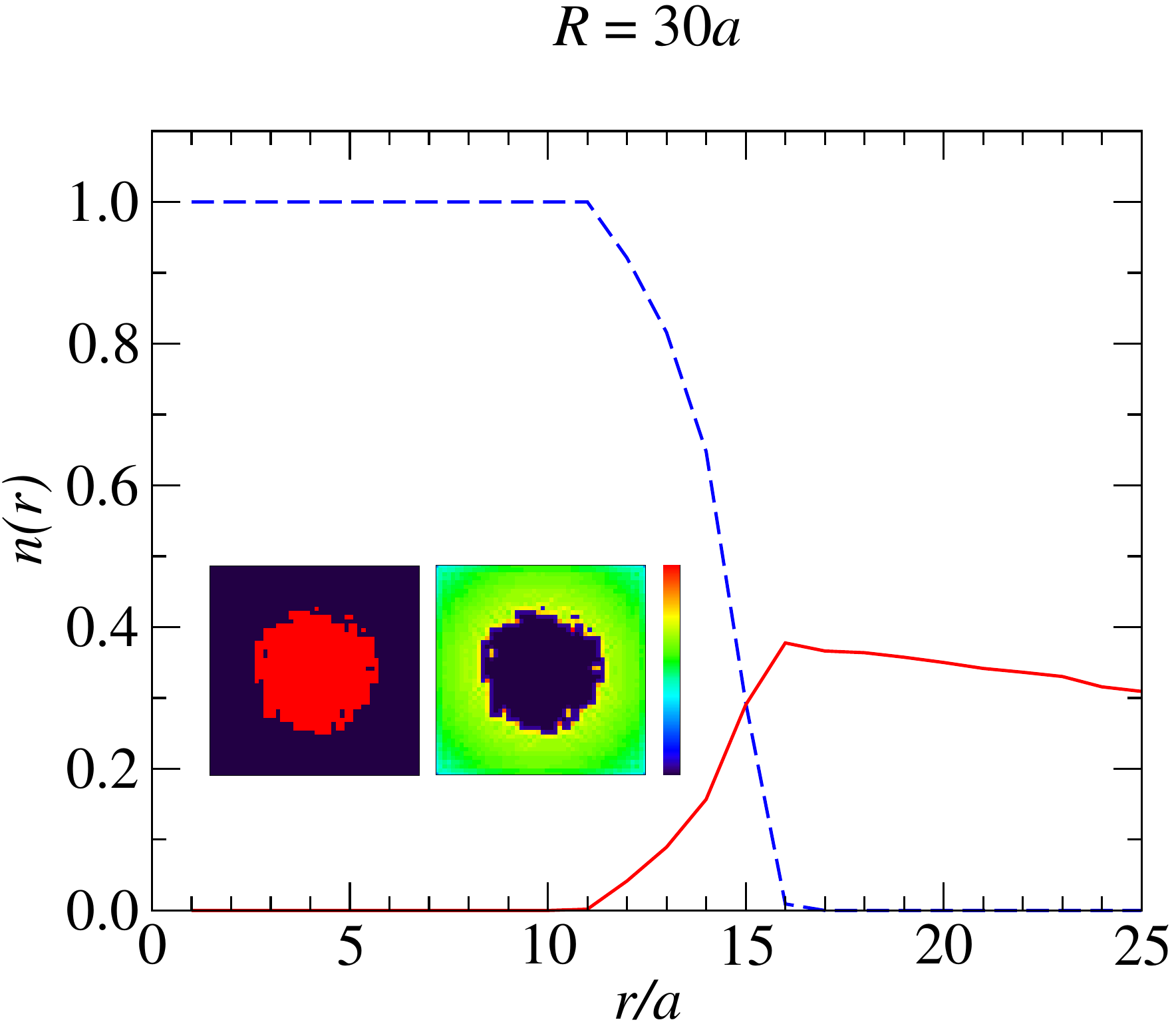}
\caption{(Color online) Density profiles of the light (solid red line) and heavy (dashed blue line) atoms
for $U=5J$ at a temperature $k_{\rm B}T=0.01 J$. The curvature of the trap for the heavy atoms is given 
by $R^{h}=30a$, whereas for the light atoms it varies from $R=12.9a$ to $R=30a$. The insets show
snapshots of real-space distributions of the heavy (left box) and light (right box) atoms.}
\label{fig1}
\end{figure*}

We ignore dynamic interactions between heavy and light atoms due to scattering in 
the course of the expansion in the time-of-flight experiments. This is a reasonable
approximation since the density of the clouds is very small once the trap and the optical lattice potentials are dropped.  Since those potentials are dropped over a finite period of time, there can be some effects as the potentials are lowered, but these tend to be smaller in these systems because the repulsive interspecies interaction keeps the light and heavy particles away from each other in their initial distributions on the lattice prior to the time-of-flight experiment.

Since in our approximation the heavy atoms are localized, we are interested in 
the momentum distribution of only the light ones. The heavy atoms, however, may
display some kind of density-wave ordering, which has been demonstrated in Ref. \onlinecite{PRL}. 
In particular, for some regimes of parameters they form a checkerboard pattern
where the heavy atoms occupy only, {\it e.g.}, black squares. In another regime,
superpositions of vertical and horizontal stripes or phase separation, with the heavy 
and light atoms occupying different regions of space have been observed.
Of course, for any nonvanishing interaction between both species of atoms the
distribution of the light atoms is up to some degree correlated with the distribution
of the heavy ones. As a result, also the light atoms are ordered, but unless the 
interaction is strong enough, the magnitude of the density-wave order is much smaller. This follows
from the fact that the light atoms, in contrast to the heavy ones, gain kinetic
energy while delocalized.

The ordering of the heavy atoms can be analyzed by use of scattering of light 
(Bragg scattering, see {\it e.g.}, Ref. \onlinecite{Bragg}). This kind of experiment is
an equivalent to neutron or X-ray diffraction for the solid state. However, due to the 
difference of lattice constants between solid state crystals and optical lattices, the 
required wavelength corresponds to that of visible light. The observation of well-defined
Bragg peaks has been used to confirm a crystalline structure formed by atoms in
an optical lattice \cite{Bragg}. Since this method allows one to determine the distance 
between atoms forming the crystalline structure, it can be applied in order to 
detect the checkerboard pattern as well. In the case of strong repulsive 
interaction between the light and heavy atoms, checkerboard squares of different
color are occupied by different kinds of atoms and the problem is similar to the detection
of antiferromagnetic (AF) order. It has recently been proposed to use Bragg 
diffraction of light to detect such an order \cite{Bragg_AFM} in a Hubbard system. In the case of AF 
order, the spin--dependent scattering is achieved by using the probe light
frequency near atomic resonance, where the interaction between light and atoms
is neither purely diffractive nor purely absorptive. In a similar way the probe light can be 
tuned to be scattered in a different way by the light and heavy atoms. As a result,
we can expect a $(\pi, \pi)$ peak in the case of a two-dimensional checkerboard pattern 
with the light scattered by the heavy atoms. Moreover, the intensity of this peak
can give information about the fraction of the system occupied by ordered atoms.
This method can also be used to detect other types of correlations. If the ``labyrinthine''
patterns obtained in Monte Carlo simulations \cite{PRL} are superpositions or mixtures 
of horizontal and vertical stripes, the Bragg scattering in this case should reveal
$(0, \pi)$ and $(\pi, 0)$ peaks.

\begin{figure*}[htb]
\includegraphics[width=0.33\textwidth]{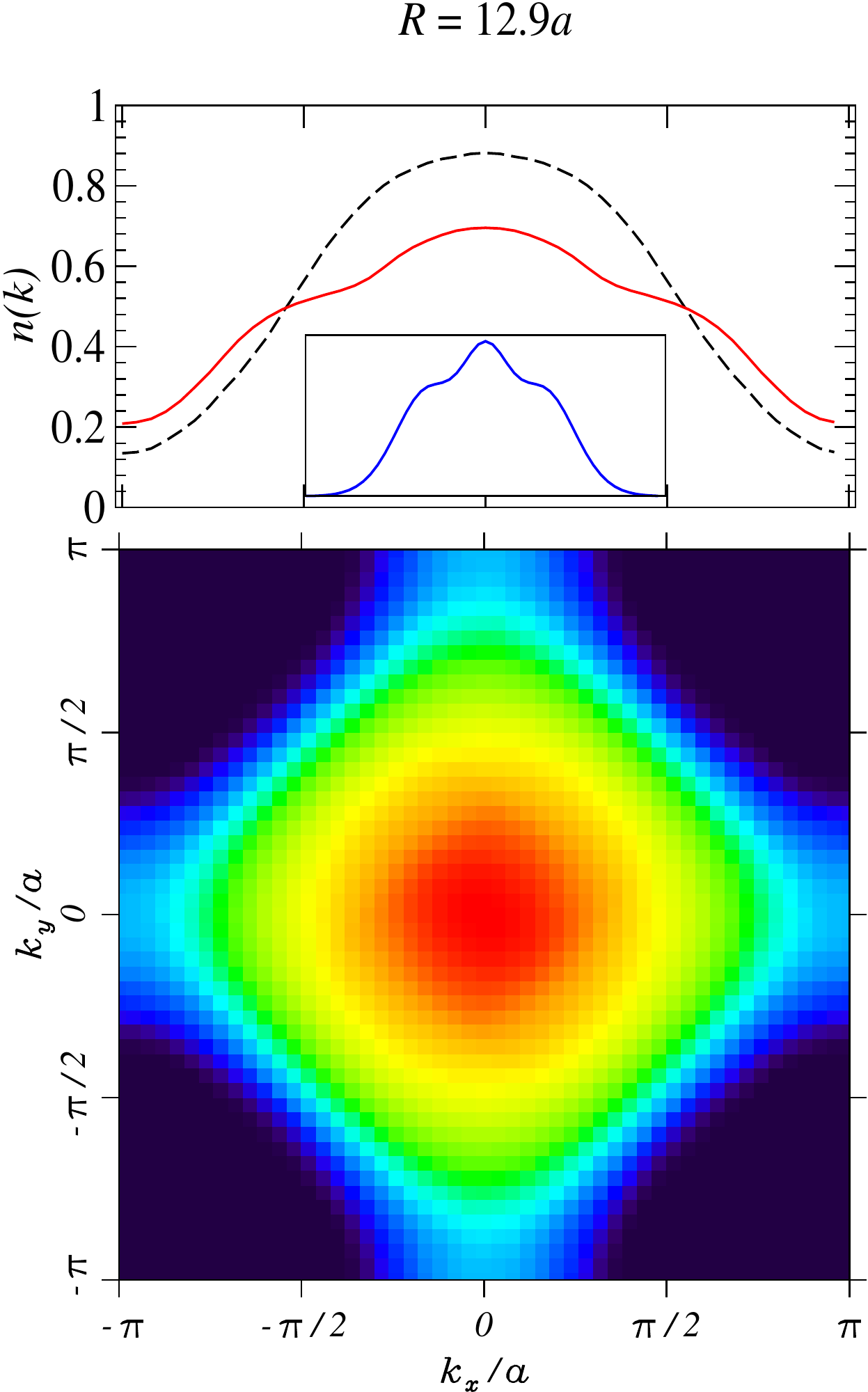}\includegraphics[width=0.33\textwidth]{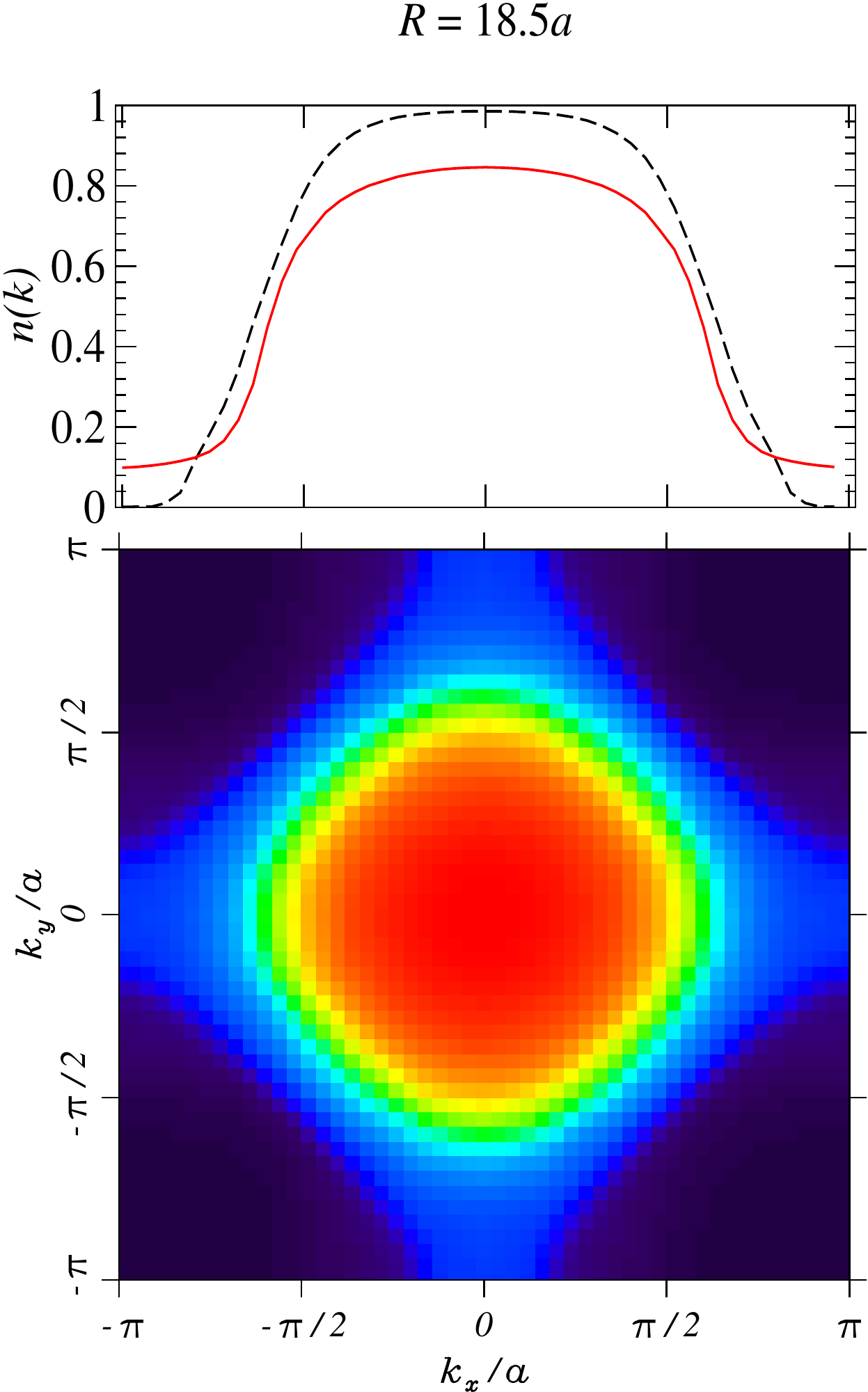}\includegraphics[width=0.33\textwidth]{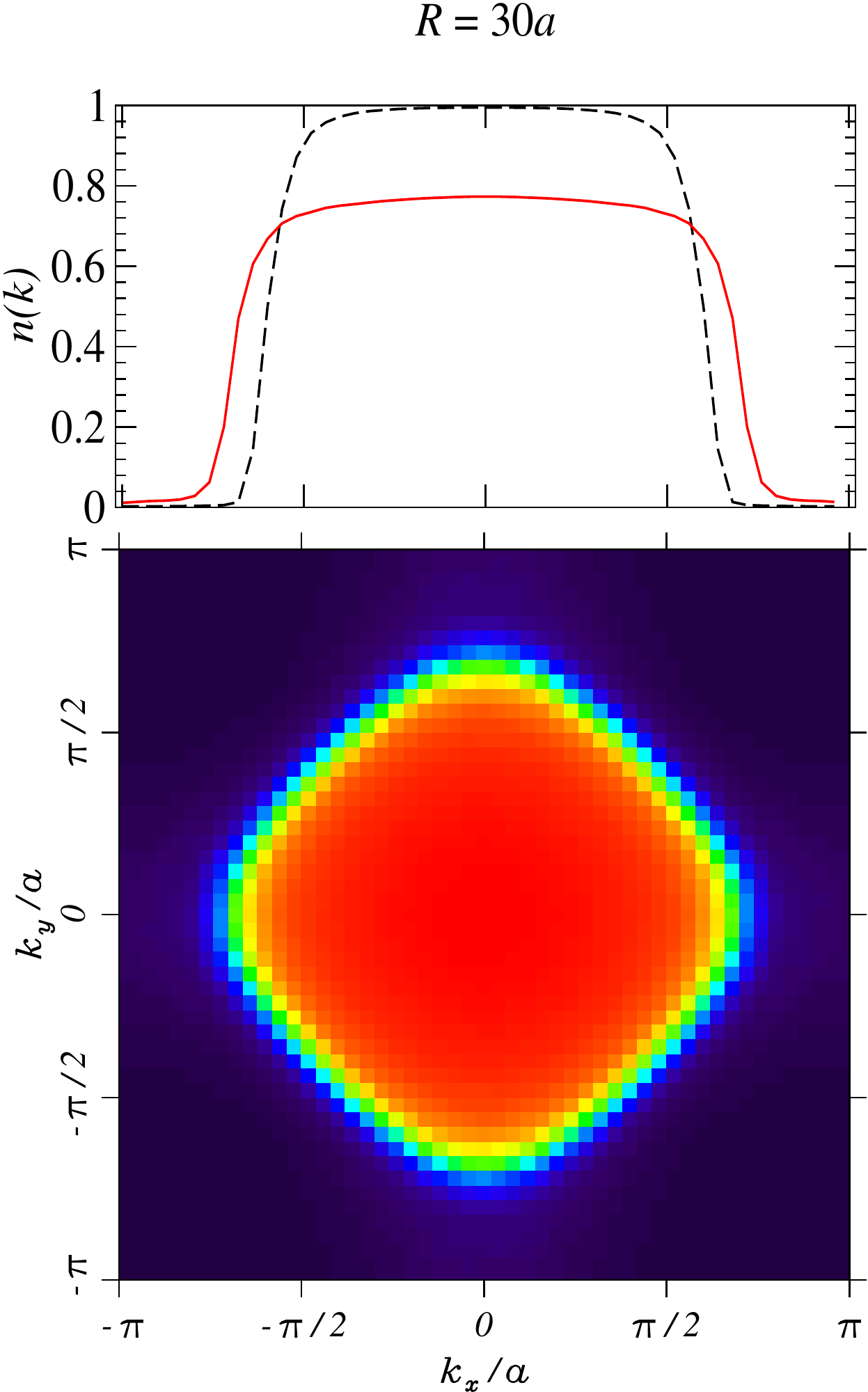}
\caption{(Color online) Momentum distribution of the light atoms for the cases shown in Fig. \ref{fig1}. The
upper row shows a comparison of cross sections along the $x$ axis for $U=5J$ (solid red line) and
noninteracting case (dashed black line). The inset in the first plot shows the momentum distribution
of the light atoms confined in a circular well with infinite walls (see text).}
\label{fig2}
\end{figure*}

The Bragg spectra integrated over the frequency gives the static structure factor 
$S({\bm k})$, a key quantity that can be expressed as
\begin{equation}
S({\bm k})=\frac{1}{N}\sum_{i} w_i e^{i{\bm k}\cdot{\bm R}_i}.
\label{bragg}
\end{equation}
In a practical realization we get a new configuration $\{w_i\}$ in each MC step and 
in each step $S({\bm k})$ is calculated. Then, $S({\bm k})$ is averaged over the entire 
MC run.

\section{Results}

Simulations have been carried out on a $50\times 50$ square lattice with 625 heavy and 625 
light atoms. In contrast to solid state simulations, there is no translational 
symmetry in the real system and therefore we use hard-wall boundary conditions at the edges.
For steep harmonic trap potentials (and at low $T$) most of the atoms are in the center of the system,
however, if the potentials are shallow the results may be affected by the
finite size of the cluster. Fig. \ref{fig1} illustrates the real-space distributions
of the light and heavy atoms for different shapes of the trap for the light atoms 
[see Eq. (\ref{trap}) for the trap potential parametrization].
The trapping potential for the heavy atoms is kept at $R^h=30a$ for all simulations. 

From Fig. \ref{fig1} one can see that if the light atoms' trap potential is sufficiently steep 
and the repulsion between the two species of atoms is relatively strong ($U=5J$), the heavy atoms
are pushed out from the center of the trap. This is an intuitive result. However, when the
trap potential for the light atoms becomes shallower they start to spread and occupy the periphery
of the cluster even before the trap curvatures are set to be equal. For identical potentials, the heavy atoms
are concentrated in the central part of the trap, surrounded by the light atoms. In such a 
configuration the light atoms can gain kinetic energy at the expense of the potential energy
in the harmonic trap. 

\subsection{Momentum distribution of the light atoms}
\begin{figure}[h]
\includegraphics[width=0.5\textwidth]{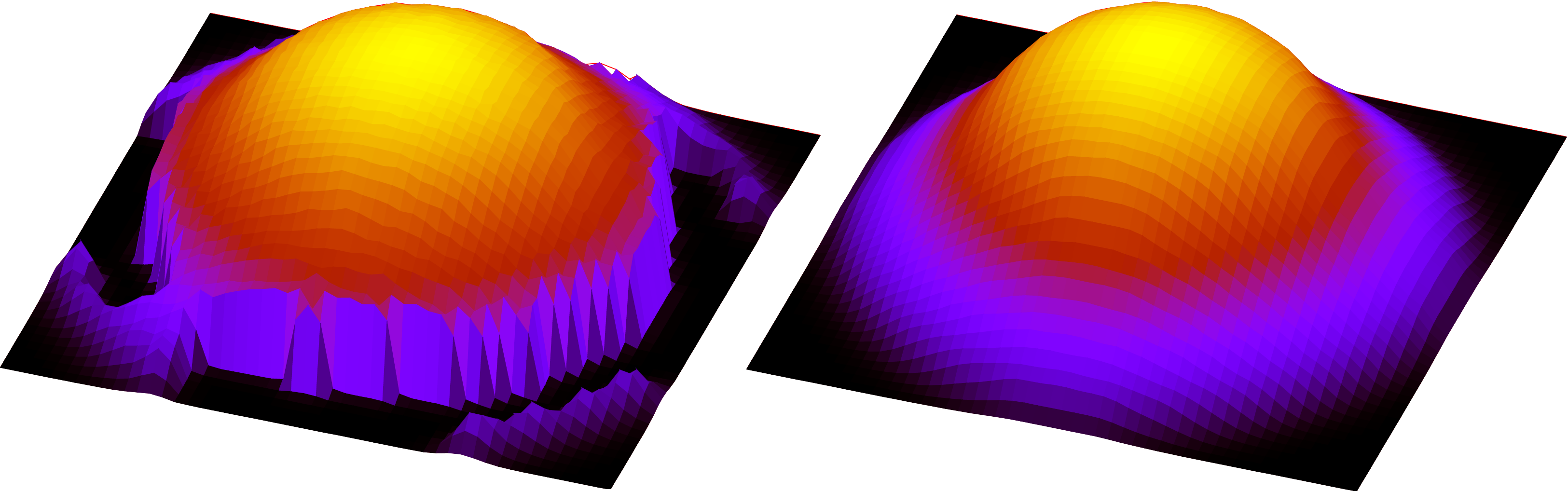}
\caption{\label{xxx}(Color online) Comparison of the real-space distribution of light 
atoms with (left graph, $U=5J$) and without (right graph) interaction with heavy atoms 
($R=12.9a$).}
\end{figure}

Fig. \ref{fig2} shows the momentum distribution of the light atoms for the same parameters 
as in Fig. \ref{fig1}. Since the strong repulsion leads to phase separation, at least
in the limiting cases, the light atoms are able to move freely within the region not occupied
by the heavy atoms. Nevertheless, one can notice that there is no sign of the Fermi surface
in the momentum distribution, which results from the inhomogeneity of the system. The upper row of panels presents a comparison of the crossections of 
the actual distribution with that of a noninteracting fermionic gas in a harmonic trap. The 
difference results from the interaction between the light and heavy atoms. Generally, for
all the shapes of the trap potential, the maximal value of the momentum distribution function at ${\bf k}=0$ of the light atoms
is reduced with respect to the noninteracting case. For a relatively steep trap potential 
for the light atoms, the interaction leads to an occurrence of additional features in the
momentum distribution [see the red (solid) curve in the upper left panel of Fig. \ref{fig2}].
The occurrence of these additional inflection points results from a further confinement of the light
atoms in an area surrounded by the heavy atoms, as can be seen in Fig. \ref{xxx}. 
In order to confirm the origin of these features, we calculated the momentum distribution of 
the light atoms in an infinite round well with a diameter equal to the inner diameter of the ring 
formed by the heavy atoms. In the resulting distribution, presented in the inset in Fig. 
\ref{fig2}, the additional peak in the center of the distribution is even more pronounced.
 
Independent of the strength of the interaction, the inhomogeneity destroys
any signs of the Fermi edge in the momentum distribution. When the trap becomes shallower,
behavior that looks like a Fermi edge appears, but this is due to the box boundary conditions at the edge of the lattice.

Even though there is no explicit signal in the momentum distribution function which shows the presence of different kinds of density wave ordering, the momentum distribution function does have a strong dependence on the appearance of phase separation, as can be seen in the different distribution functions in Fig.\ref{fig2}.  For example, it is known that in a homogeneous lattice, the momentum distribution function must decrease below the noninteracting value for small momentum and increase above the noninteracting value for large momentum~\cite{ueltschi} when the system phase separates at low temperature.  This behavior is clearly seen in all of the data, and most likely is arising from different forms of quantum confinement effects associated with the phase separation.  Unfortunately, it is not easy to disentangle this phase separation effect from the effect of the trap, which has a similar effect on the momentum distribution function, except in the case where the heavy particles surround the light ones and confine them with a sharp boundary.  In that case, the phase separation effect causes a ``dimple'' in the momentum distribution function near ${\bm k}=0$.

\begin{figure}
a) \includegraphics[width=0.3\textwidth]{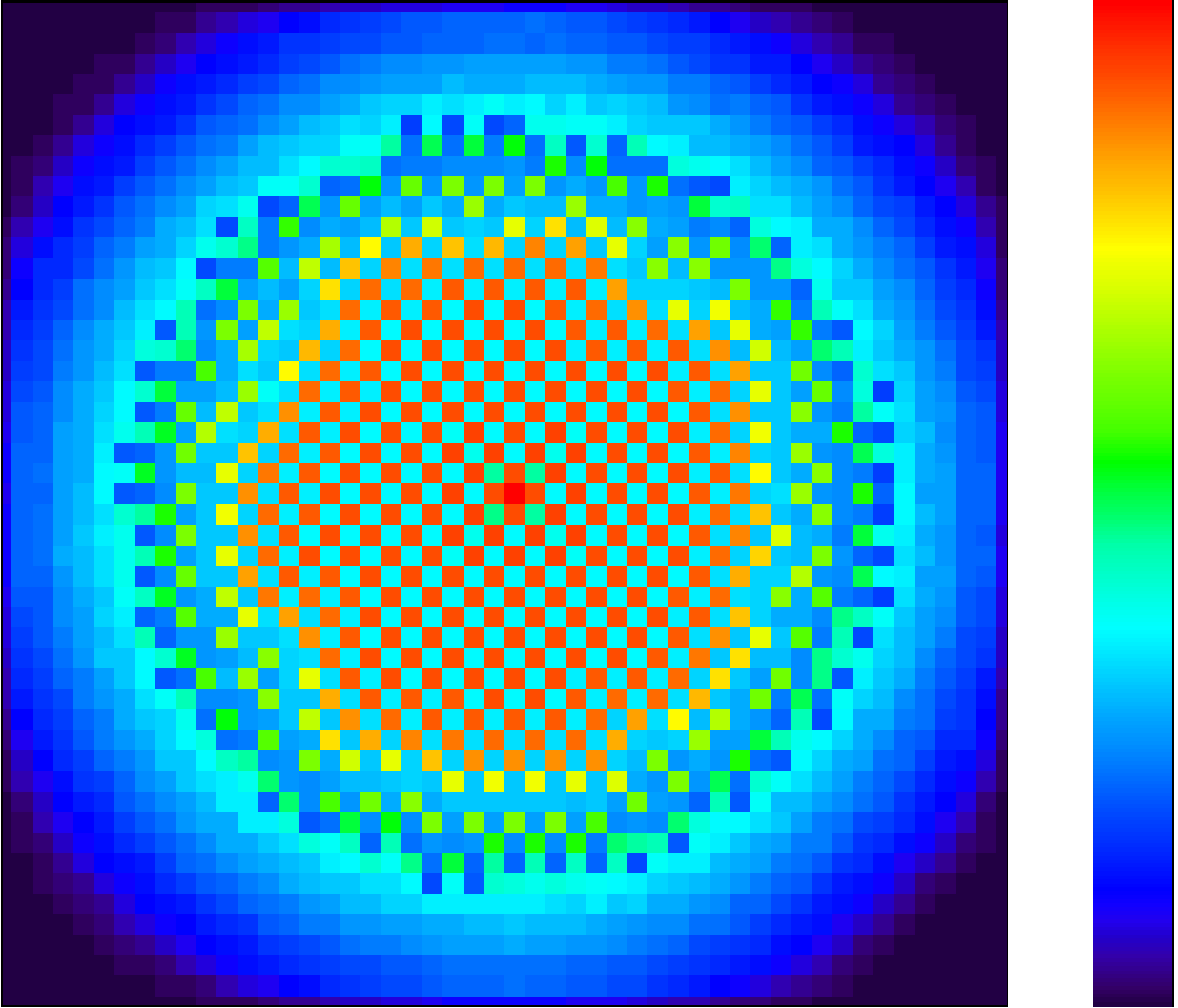}\vspace*{2mm}\\
b) \includegraphics[width=0.3\textwidth]{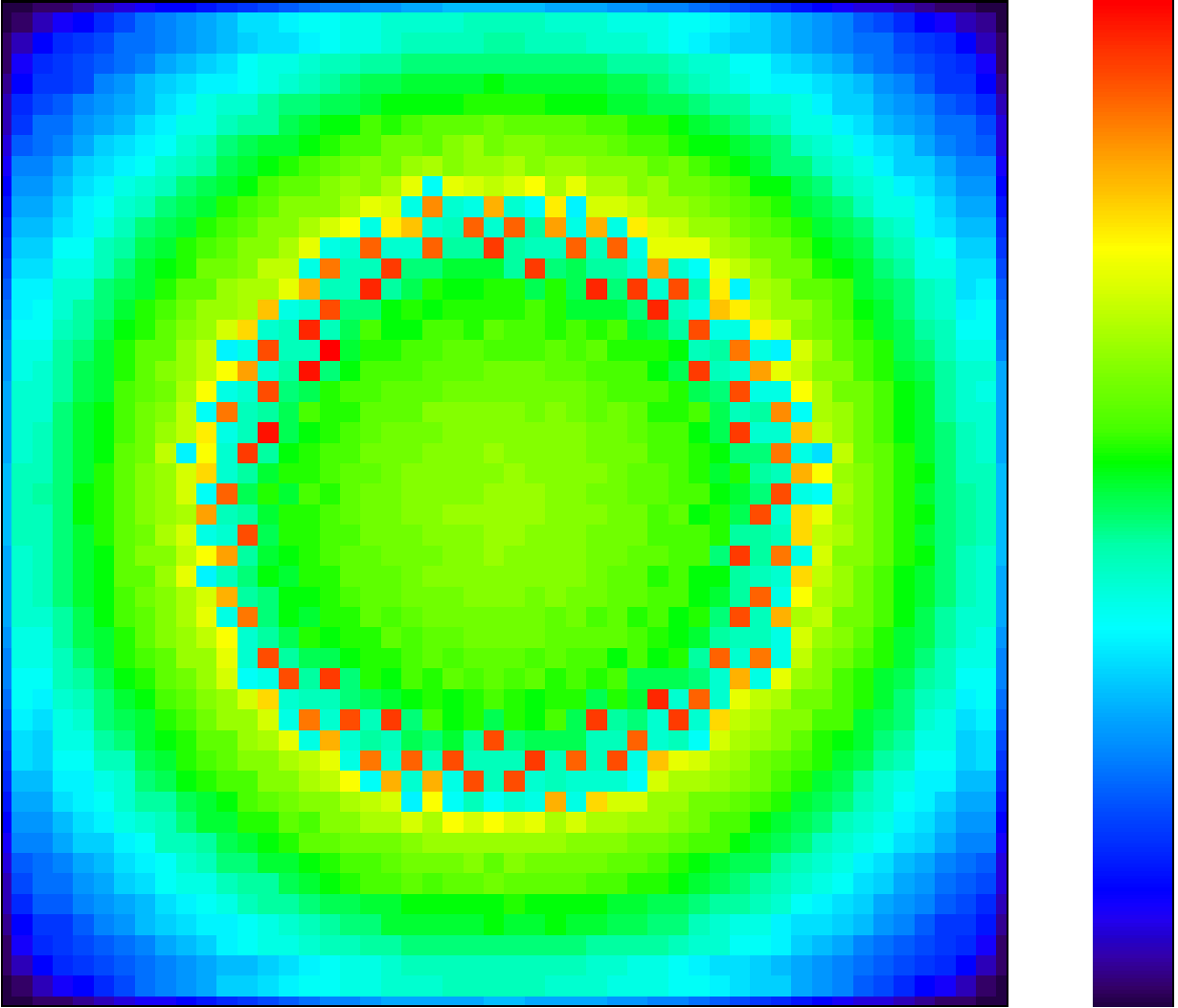}
\caption{(Color online) Real-space density distribution of the light atoms for $U=J$ and $k_{\rm B}T=0.0001J$. The 
upper panel (a) corresponds to the trapping potential with $R=12.9a$ and the lower (b) to $R=17a$.}
\label{fig3}
\end{figure}

\begin{figure*}
\includegraphics[width=0.32\textwidth]{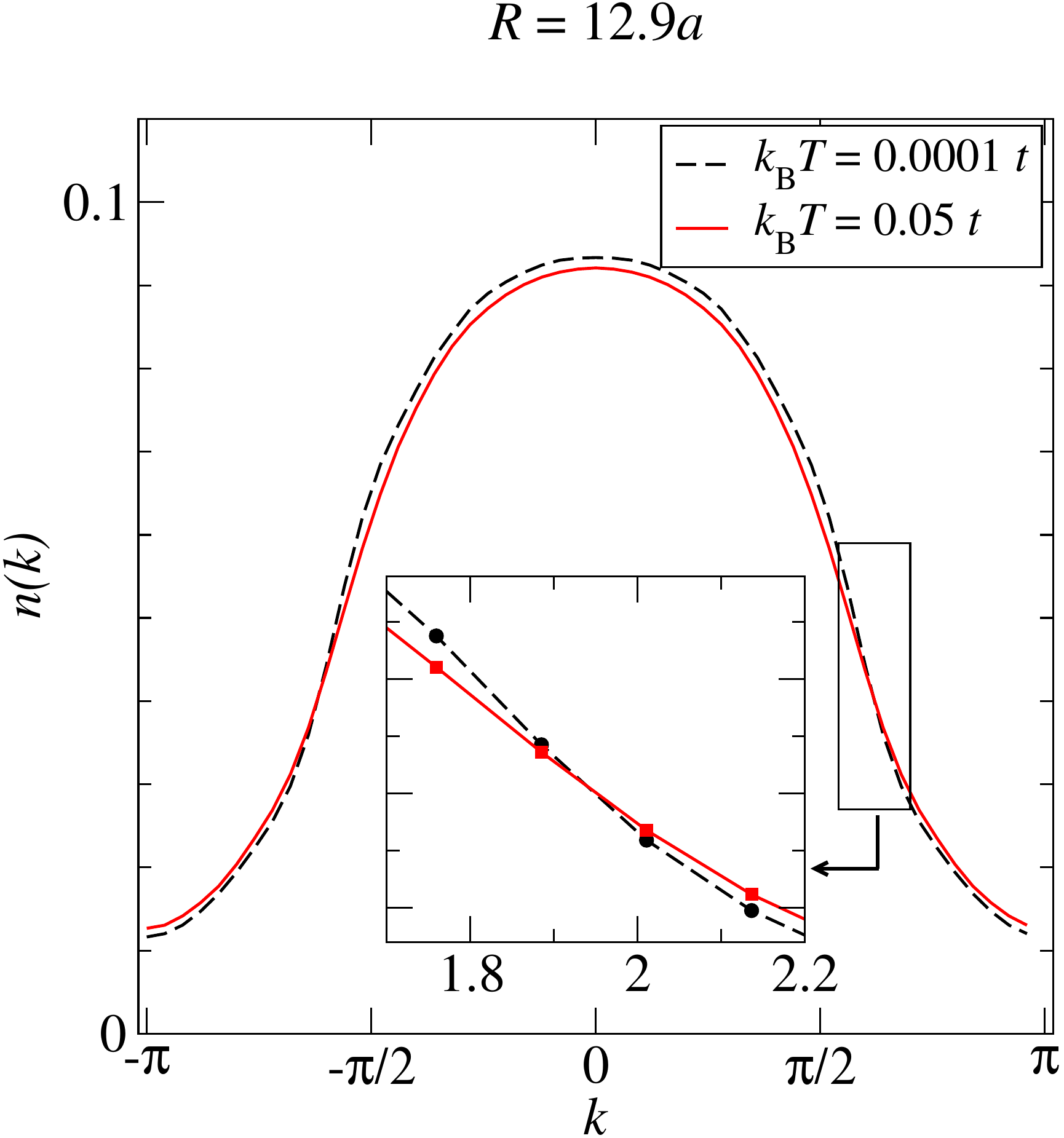}\includegraphics[width=0.32\textwidth]{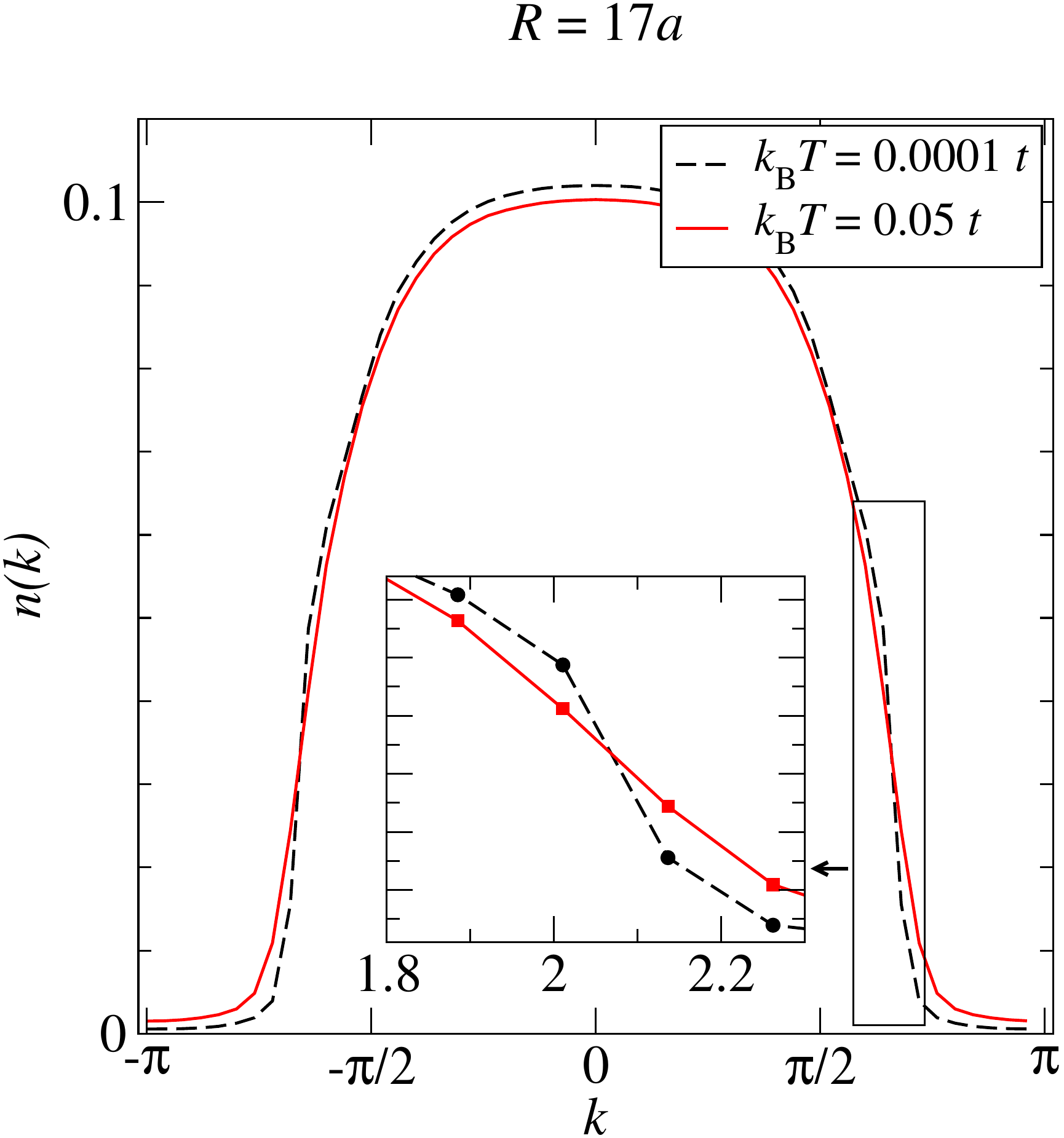}\\
\includegraphics[width=0.32\textwidth]{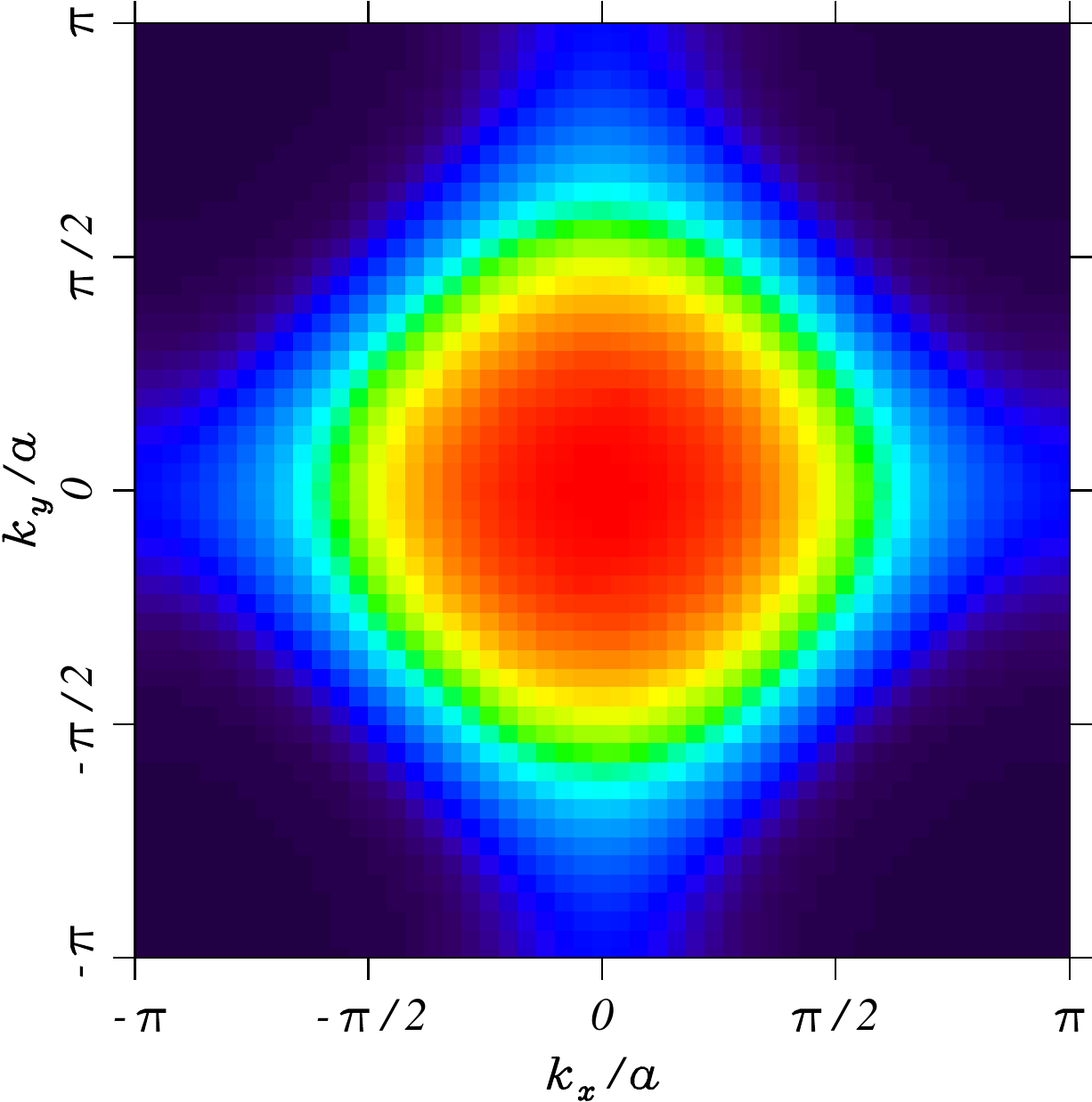}\includegraphics[width=0.32\textwidth]{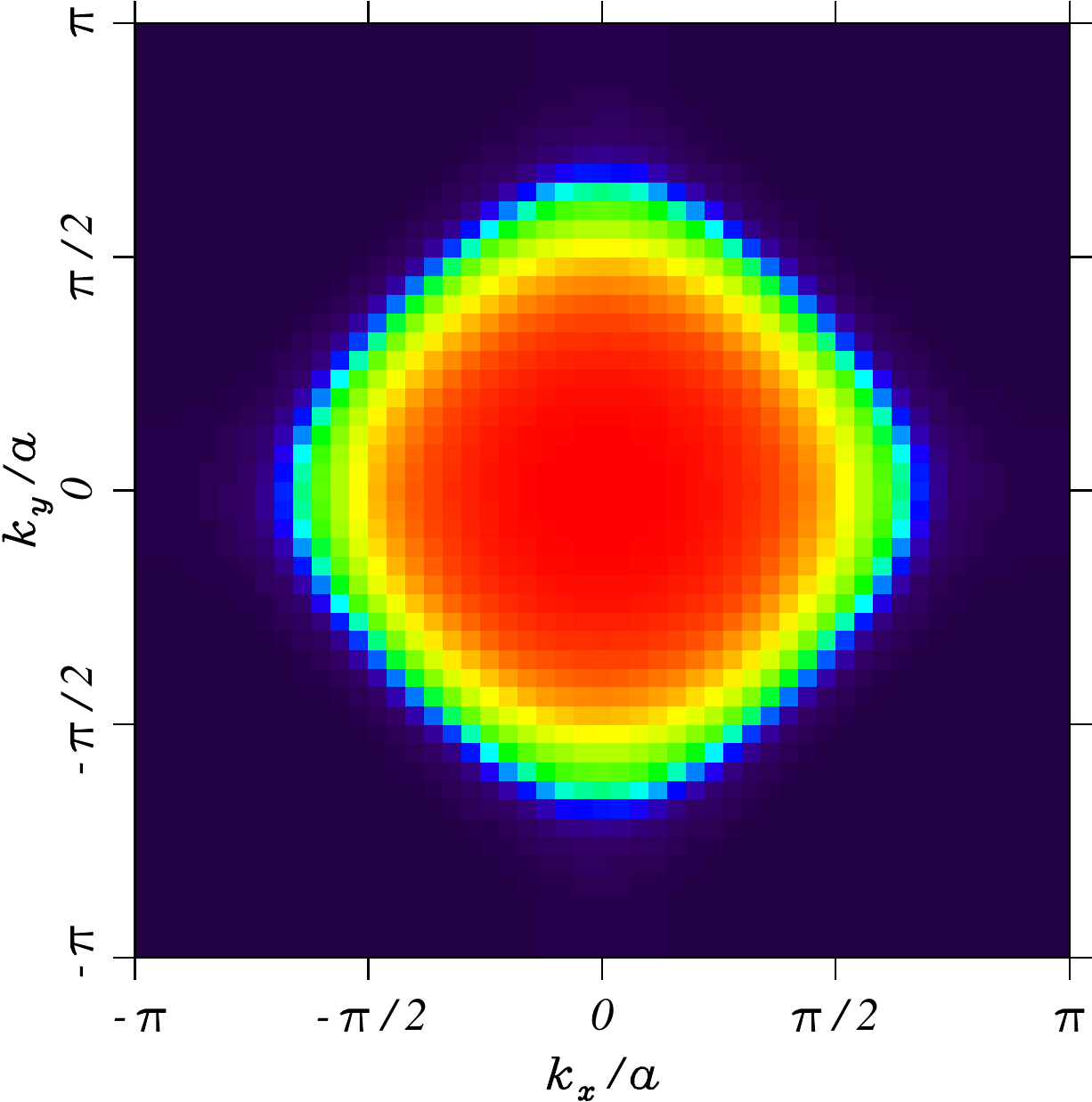}\\
\caption{(Color online) Momentum distribution  of the light atoms for the cases shown in Fig. \ref{fig3}. The 
upper row shows a comparison of cross sections along the $x$ axis for $k_{\rm B}T=0.0001J$ (dashed black line) 
and $k_{\rm B}T=0.005J$ (solid red line). The insets blows up the momentum distribution for atoms with energies
close to the Fermi energy. The momentum distributions shown in the lower row were determined at
$k_{\rm B}T=10^{-4}J$.}
\label{fig4}
\end{figure*}

If the interspecies repulsion is too weak to lead to phase separation, a checkerboard configuration may be 
formed. Fig. \ref{fig3} shows examples of real-space configurations of the light atoms for $U=J$.
It turns out that in this case the momentum distribution is hardly affected by the interaction. Fig.
\ref{fig4} presents a comparison of the momentum distribution at temperatures above and below the
temperature, at which a checkerboard pattern is formed. When the temperature is lowered, the momentum 
distribution is enhanced for small ${\bm k}$, though the difference much smaller. Moreover, formation 
of regions with density wave order does not affect the distribution in a significant way.

\subsection{Heavy atom configurations}
\begin{figure}
\includegraphics[width=0.23\textwidth]{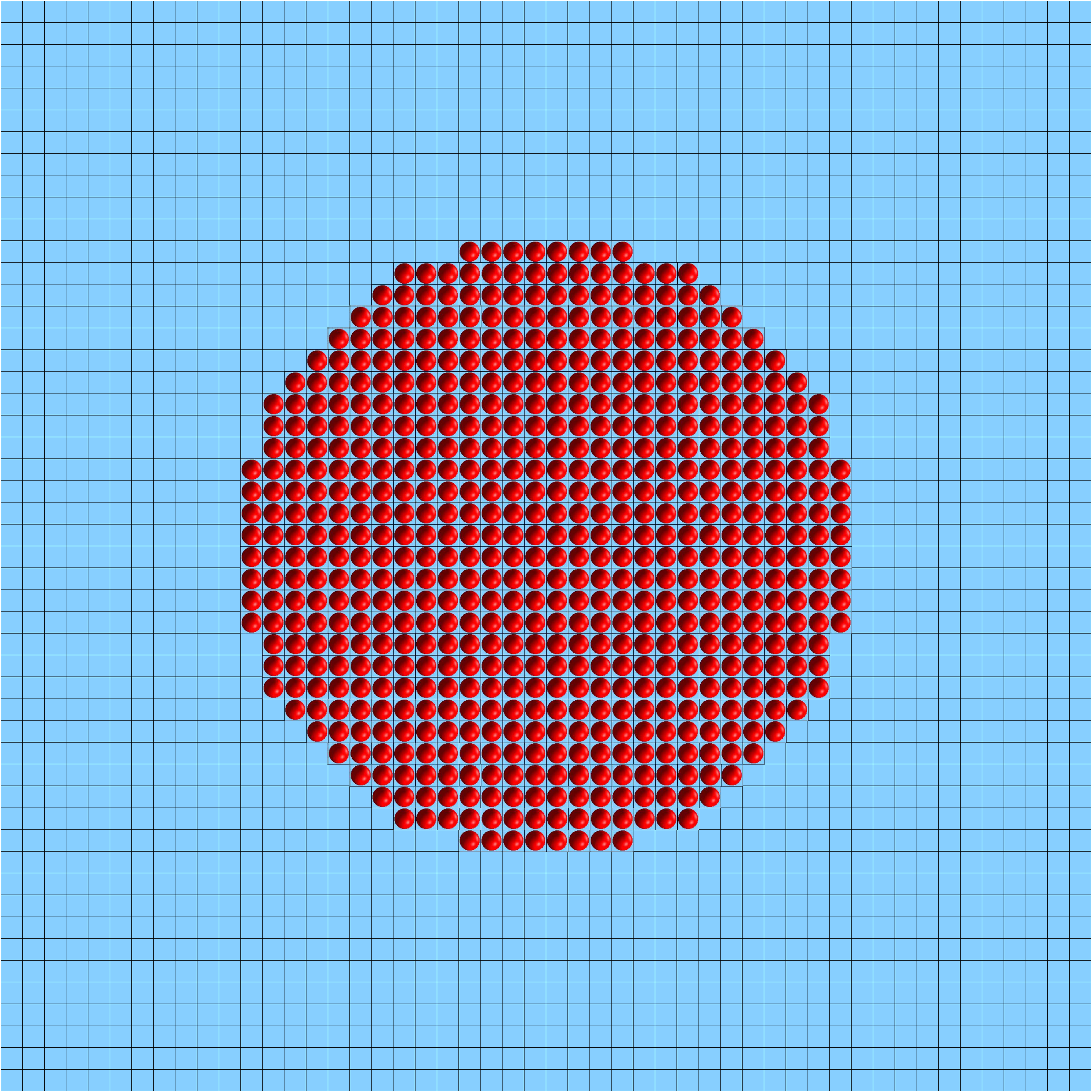}\hspace*{2mm}\includegraphics[width=0.25\textwidth]{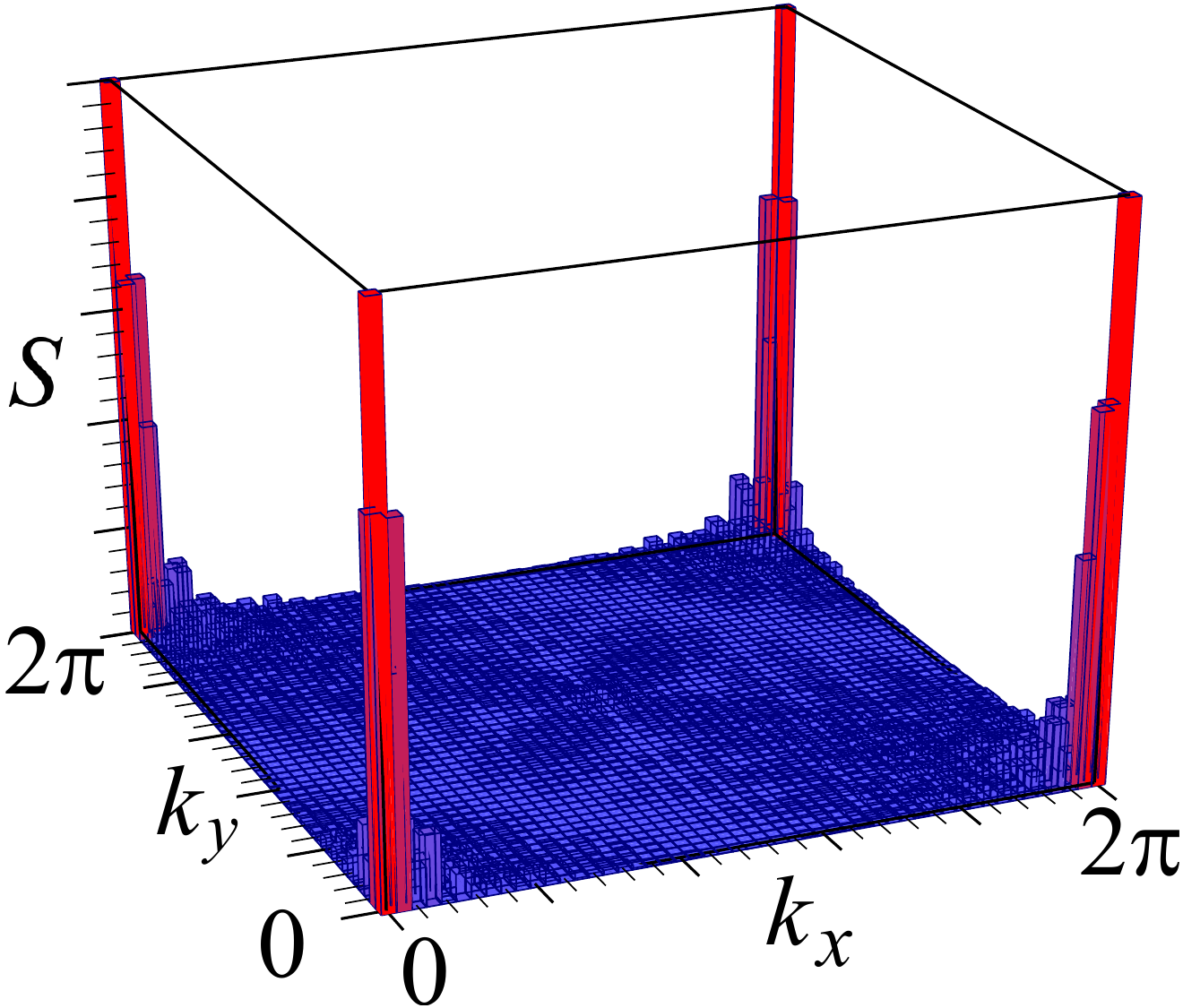}
\caption{(Color online) Left panel: ground state real-space configuration of the heavy atoms in a harmonic trap.
Right panel: corresponding structure factor.\label{circle}}
\end{figure}

Since the heavy atoms are localized in the proposed approach, one cannot analyze their momentum distribution.
Instead, we determine structure factors defined by Eq. \ref{bragg}. In particular, we are interested 
in how different density-wave-ordered patterns are reflected in the structure factor. It is well known from solid state physics that 
different orderings produce characteristic Bragg spectra. However, in the case of cold atoms the spectra are
additionally affected by the confining potential that keeps the atoms inside the trap. Neglecting the interspecies 
interaction that may lead to phase separation or ordering, at low temperature, the heavy atoms occupy the 
bottom of the trap forming a circular region. Fourier transformation of such a configuration consists of a 
finite-width peak at position ${\bm k}=(0,0)$. Plotted in a region $(0,2\pi)\times(0,2\pi)$, the peak splits into four quarters which are visible in each corner [at ${\bm k}=(0,0),\ (0,2\pi),\ (2\pi,0)$ and $(2\pi,2\pi)$]. 
The peak itself simply results from a finite number of heavy atoms, since according to Eq.~(\ref{bragg}),
$S({\bm 0})$ is equal to the average  concentration of the heavy atoms. The gathering of atoms 
in the center of the trap broadens it, as can be seen in Fig. \ref{circle}. 
When temperature or interaction moves the atoms to more peripheral areas, the peak width shrinks, finally 
taking on a delta-function-like form for a random distribution of atoms. It changes the spectral weight around 
${\bm k}=(0,0)$ (and equivalent points), affecting the magnitude of the peaks representing ordered phases.
This will be discussed in more detail below. Since we are mainly interested in patterns formed by the heavy atoms, we will 
neglect the ${\bm k}=(0,0)$ peak and focus on the remainder of the spectrum. Nevertheless, even in the presence 
of ordering, this peak is still the most pronounced feature of the spectrum. 

Most of the results have been obtained at finite temperatures by means of the Monte Carlo method. These results
can also be compared with zero-temperature local density approximation (LDA) results, which we now describe.

\subsubsection{Zero temperature results: LDA}

\begin{figure*}
\begin{center}
A)\ \ \includegraphics[width=2.75cm]{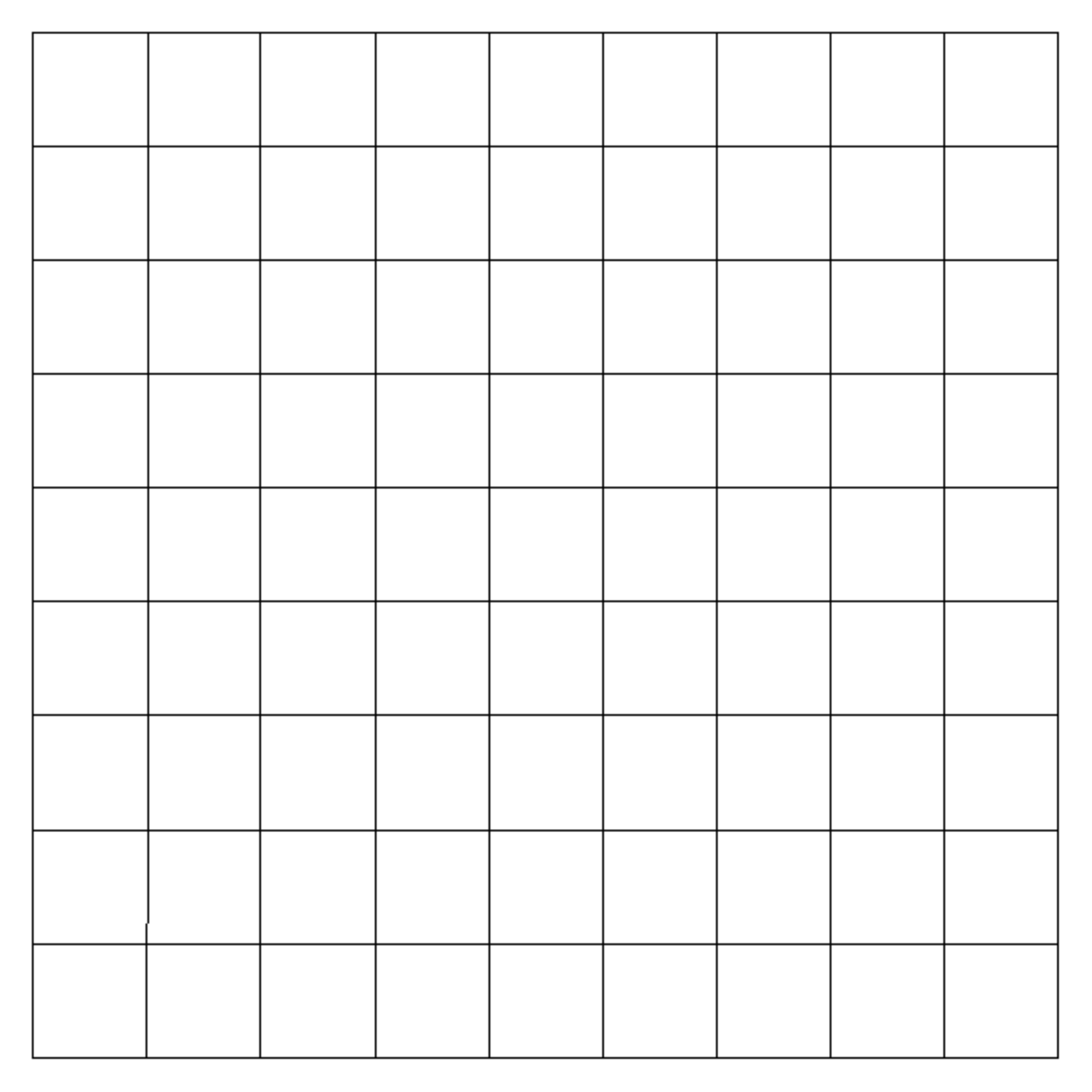}\hspace*{2mm}
B)\ \ \includegraphics[width=2.75cm]{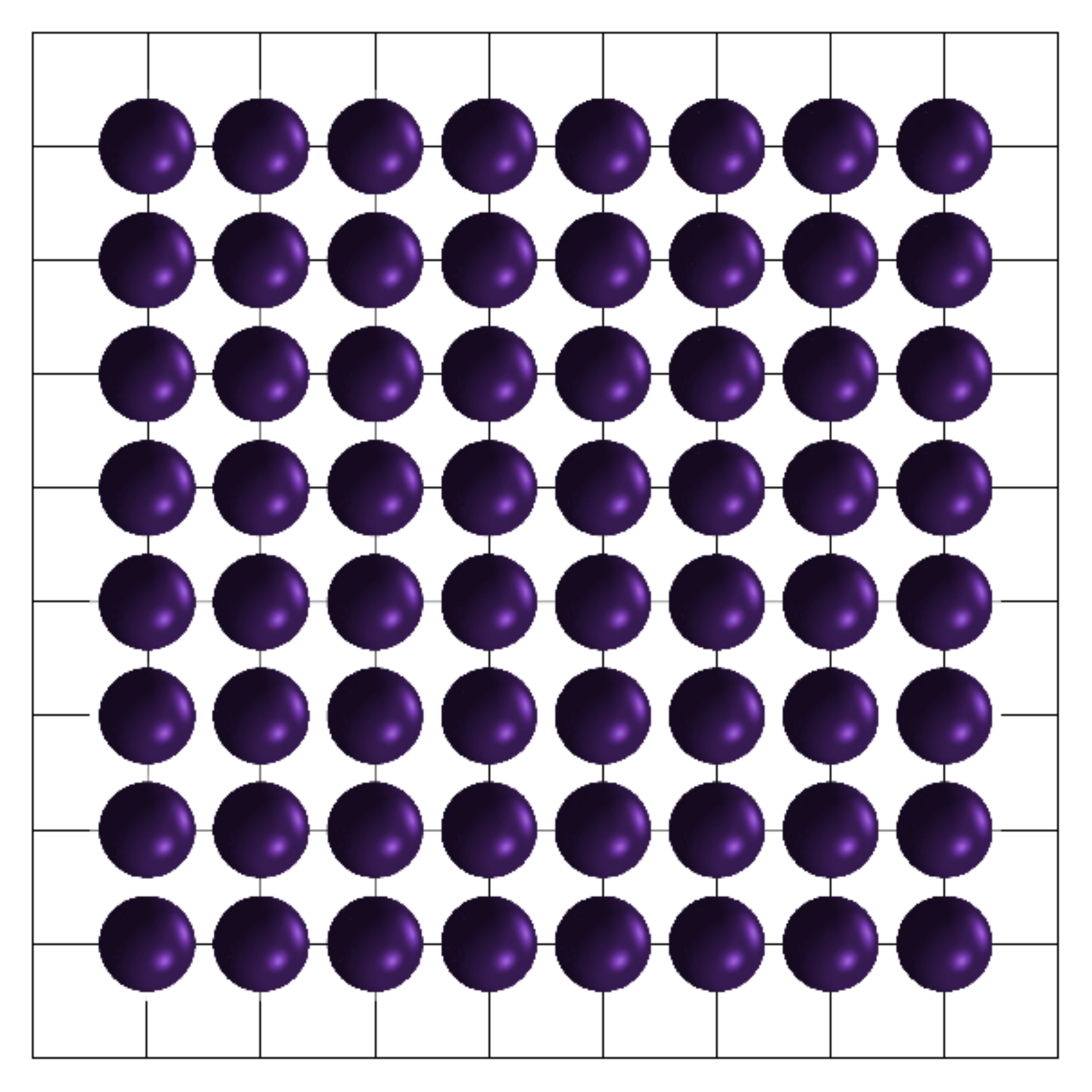}\hspace*{2mm}
C)\ \ \includegraphics[width=2.75cm]{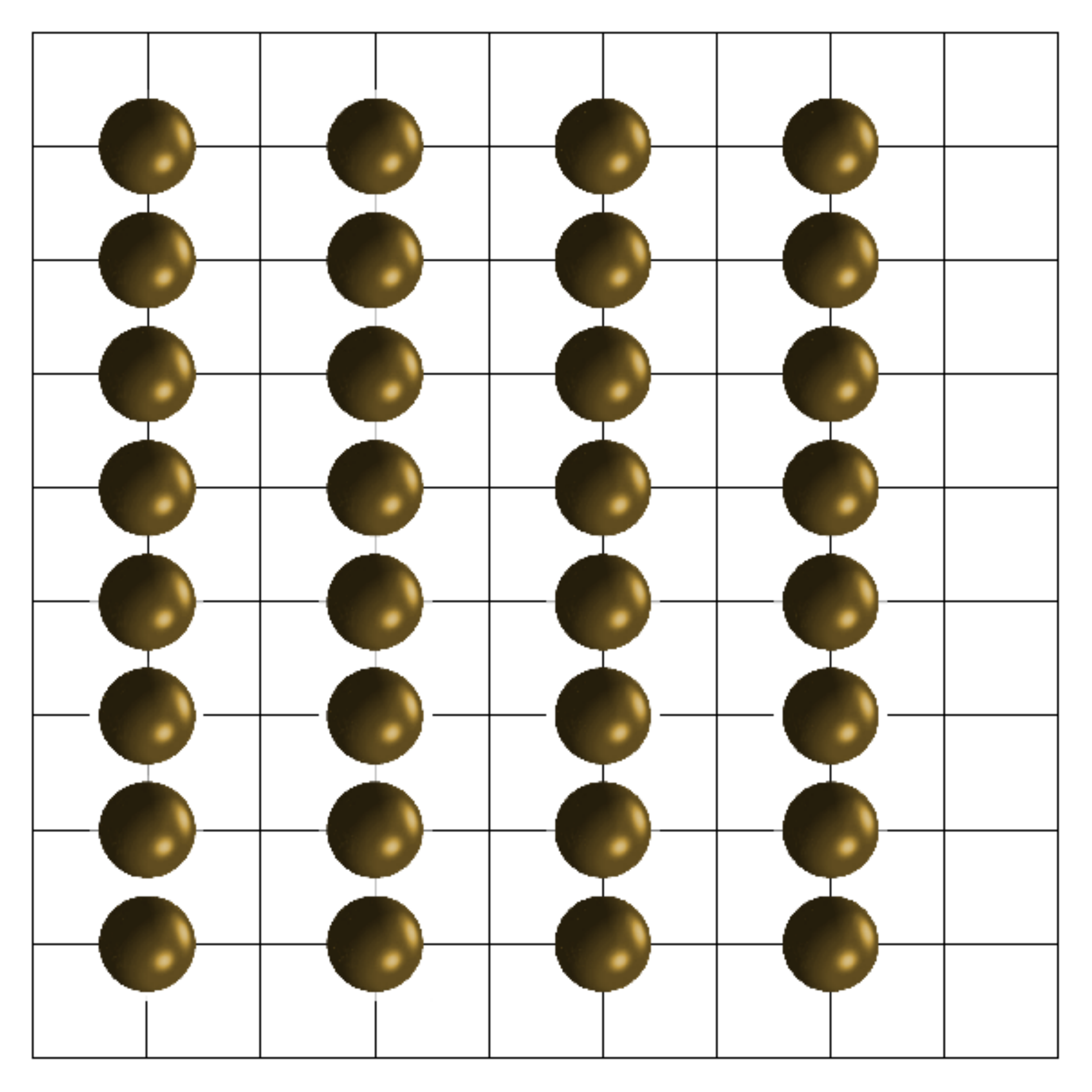}\hspace*{2mm}
D)\ \ \includegraphics[width=2.75cm]{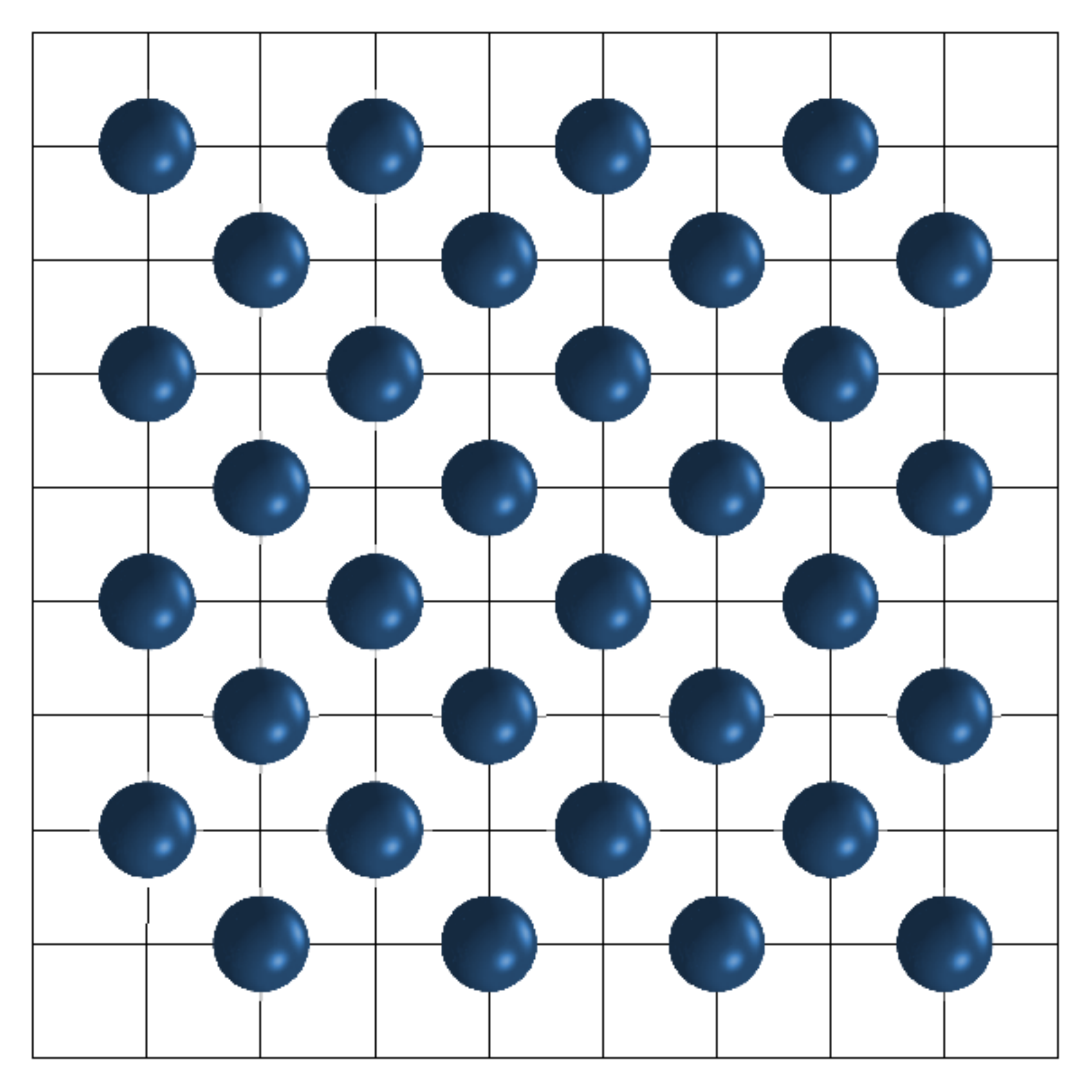}\hspace*{2mm}
E)\ \ \includegraphics[width=2.75cm]{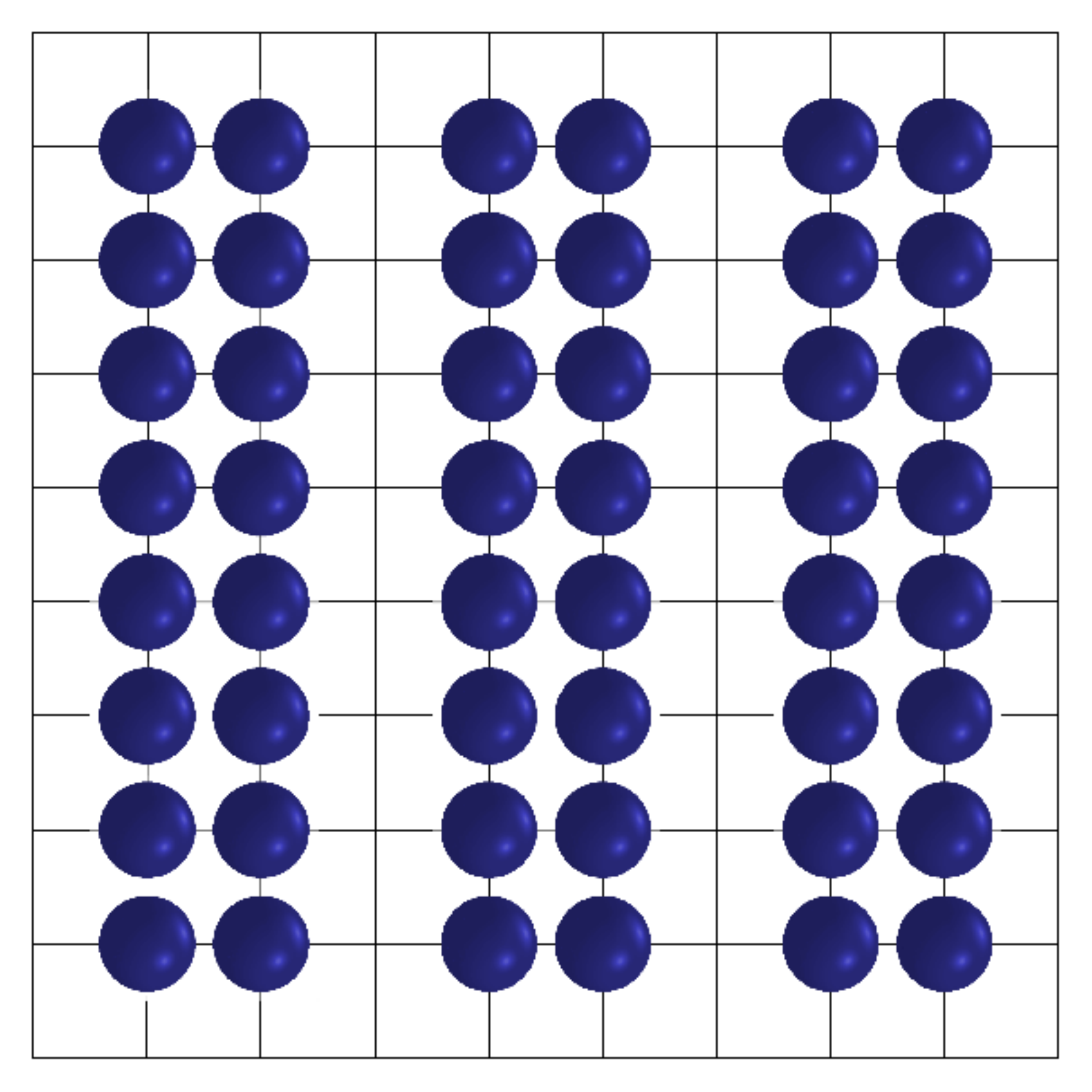}\vspace*{5mm}\\
F)\ \ \includegraphics[width=2.75cm]{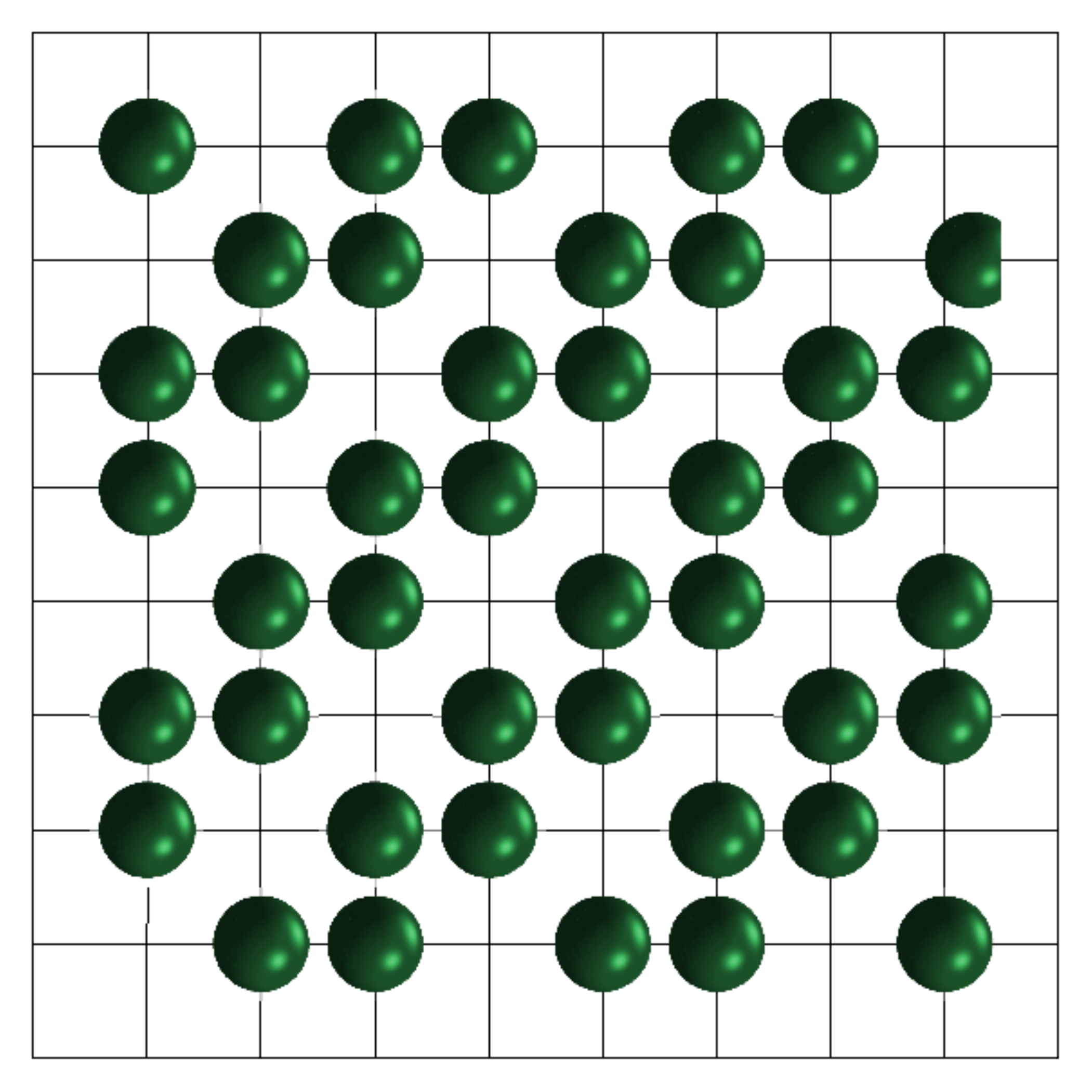}\hspace*{2mm}
G)\ \ \includegraphics[width=2.75cm]{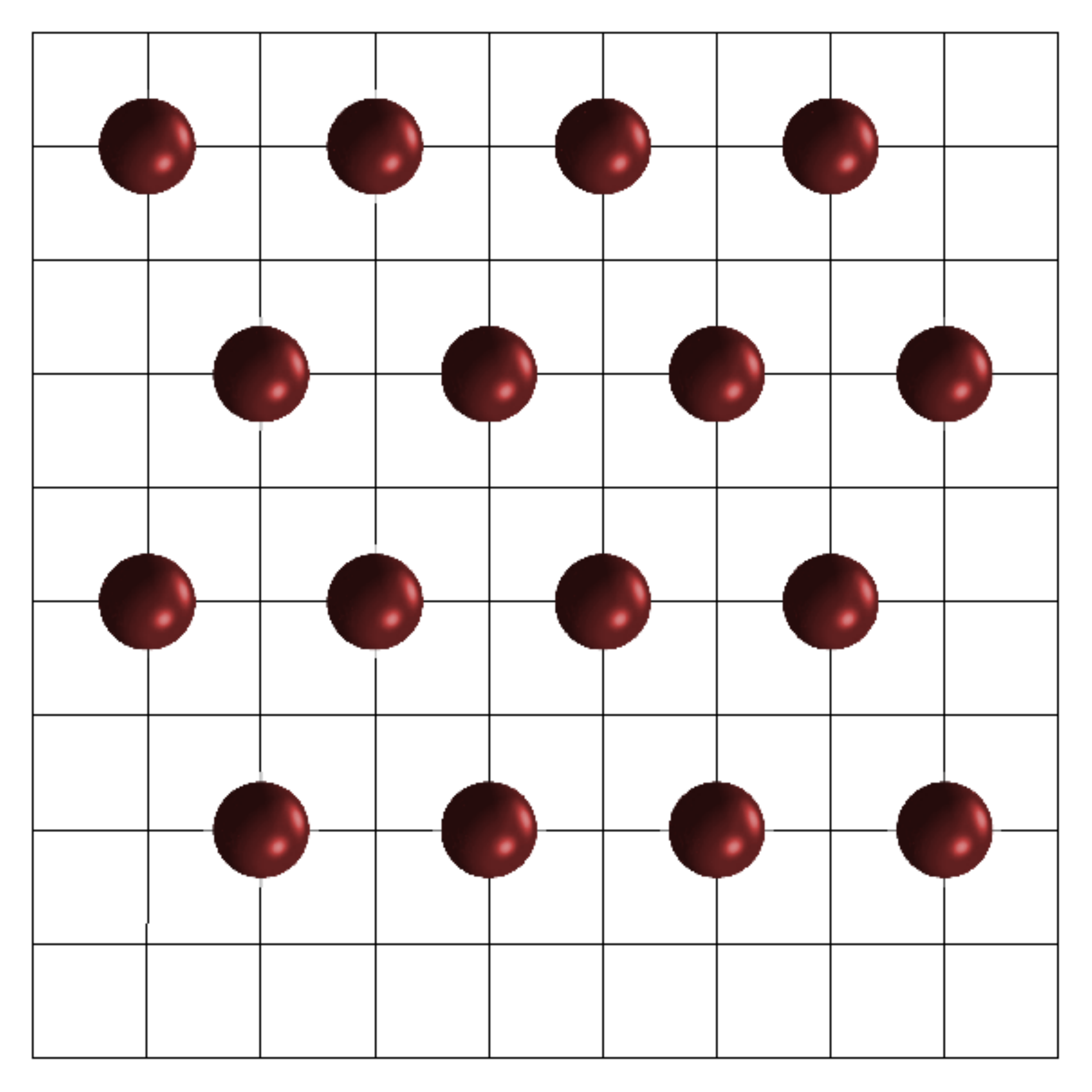}\hspace*{2mm}
H)\ \ \includegraphics[width=2.75cm]{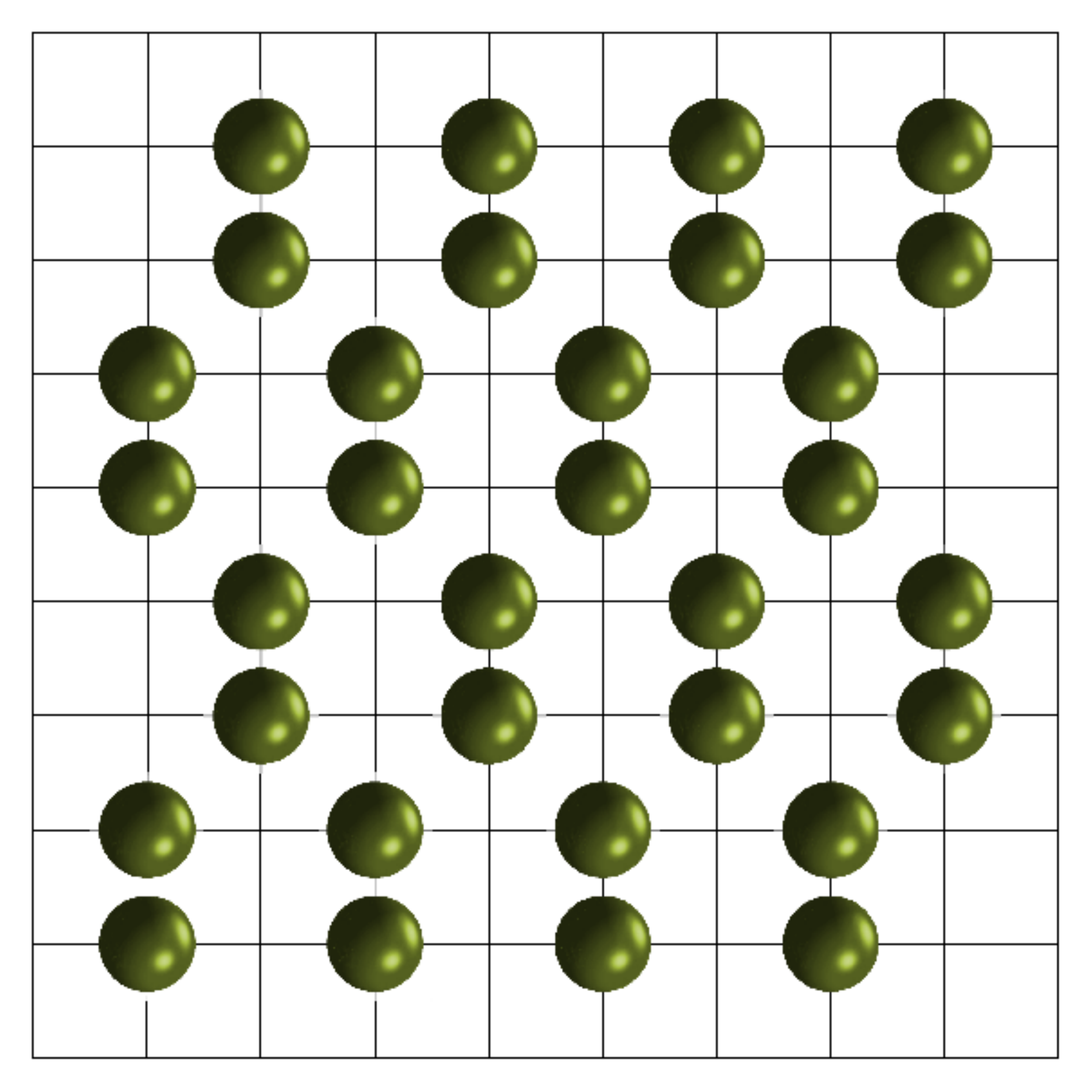}\hspace*{2mm}
I)\ \ \includegraphics[width=2.75cm]{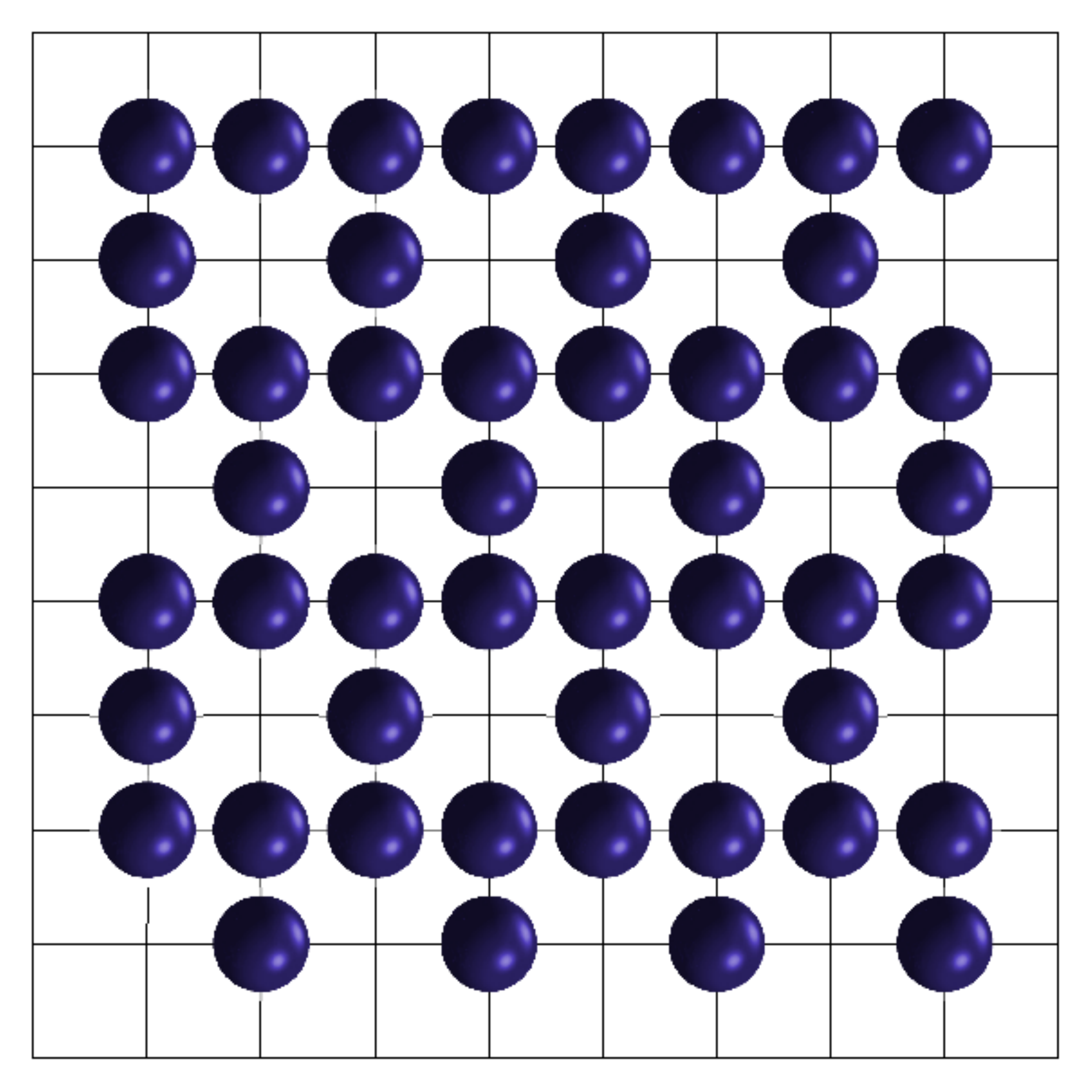}
\end{center}
\caption{\label{configurations}(Color online) All the configurations of the heavy atoms that have 
been taken into account in the LDA calculations. The letters correspond to different regions 
presented in Figs. \ref{r12_9}A and \ref{r18_5}A.} 
\end{figure*}

\begin{figure*}[htb]
\includegraphics[width=\textwidth]{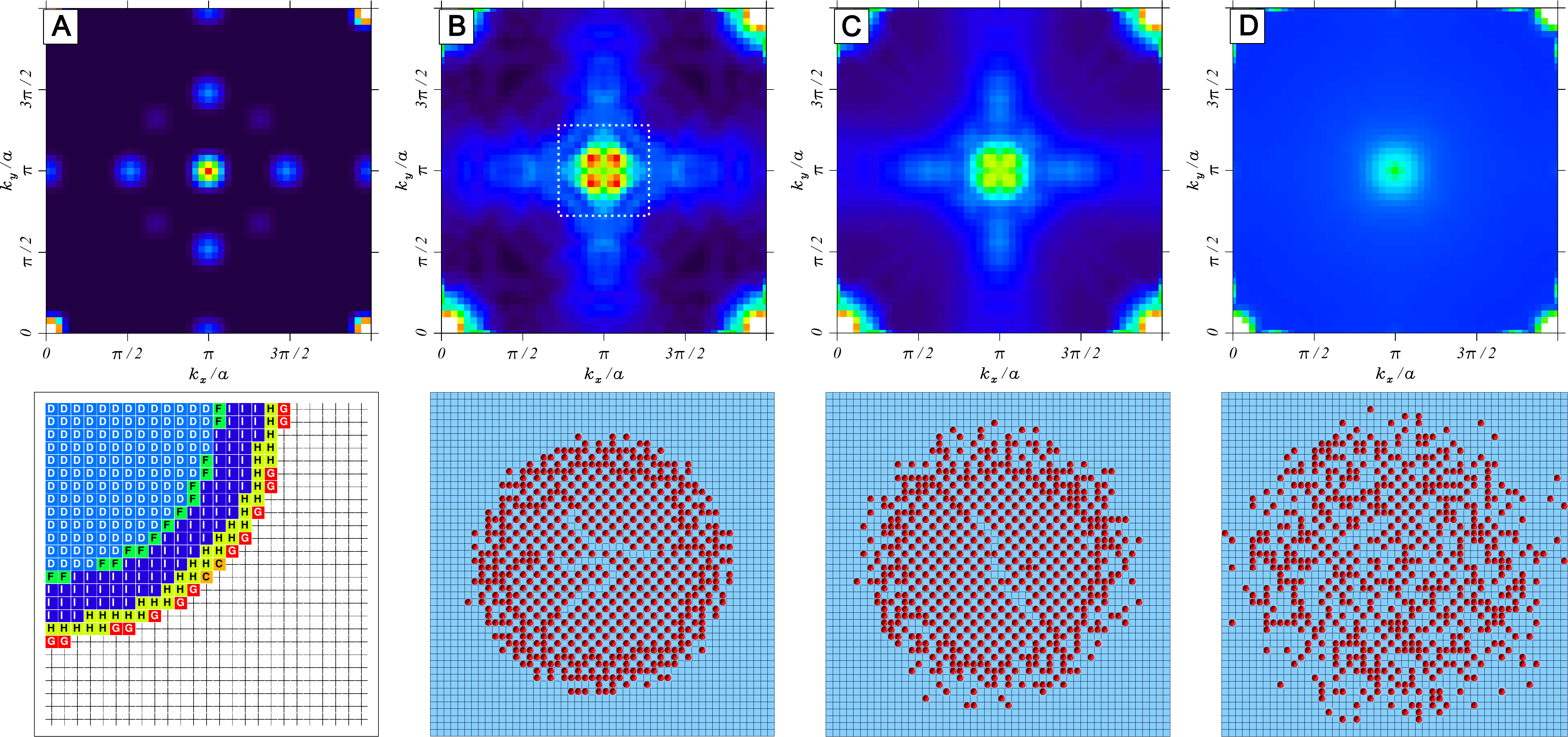}
\caption{\label{r12_9}(Color online) Upper row:  Fourier transforms of the spatial distributions of the heavy 
atoms; Lower row: corresponding spatial distributions for $U=J$. Figure A shows a 
transformation of a configuration obtained within the LDA. In the figure presenting the corresponding 
real-space configuration regions occupied by different phases are filled with different colors and 
marked by letters. Each letter stands for one configuration and the configurations are shown
in Fig. \ref{configurations}. Figures B, C, and D correspond to results at finite temperature, 
$k_{\rm B}T = 0.005J,\ 0.01J,$ and $0.05J$, respectively.}
\end{figure*}

We construct the LDA at $T=0$ from the homogeneous, grand canonical phase diagram, 
where the ground state phases are given as a function of the chemical potentials of heavy and light atoms~\cite{rom}. 
The procedure is as follows. For each lattice site, we determine the local chemical potentials by subtracting the 
trap potential at that lattice site from a trial global chemical potential. The global chemical potential is then
adjusted to produce the correct total number of heavy and light atoms in the trap. Next, using the local chemical 
potentials, we map out the homogeneous phase diagram for each site within the trap.
Two sets of model parameters that give nontrivial configurations have been analyzed: $R=12.9a,\ U=J$ and
$R=18.5a,\ U=5J$. In the former case, a checkerboard-type configuration is formed in the center of the trap, where
the heavy atoms occupy, let us say, the black squares, and the light atoms are primarily on the white ones. The 
central part is surrounded by rings of different phases. In the latter case, the center of the trap is occupied 
by light atoms, while the heavy ones form a relatively thick ring composed mainly of various stripe phases.
The spatial distributions have been obtained taking into account all possible periodic phases with unit cells
consisting of no more than 4 lattice sites (in all possible shapes). The candidate phases are presented in Fig. 
\ref{configurations}.
The zero-temperature LDA configurations are presented in the lower panels in Figs. \ref{r12_9}A and \ref{r18_5}A, 
where different letters correspond to different phases marked with the same identifying letters as in Fig. \ref{configurations}.
The Bragg spectra presented in the upper panels in Figs. \ref{r12_9}A and \ref{r18_5}A are calculated  
in the following way: first, the positions of the $\delta$-function-type peaks are determined from the Fourier transforms of the 
configurations depicted in Fig. \ref{configurations}; then the relative spectral weight proportional to the fraction of the 
trap occupied by the corresponding phase is assigned to the given peaks. Finally, the peaks are slightly broadened to 
make the presentation more clear. This is necessary because the LDA does not know about the finite size of the system and hence always displays perfect delta-function peaks.

\begin{figure}[htb]
\includegraphics[width=0.4\textwidth]{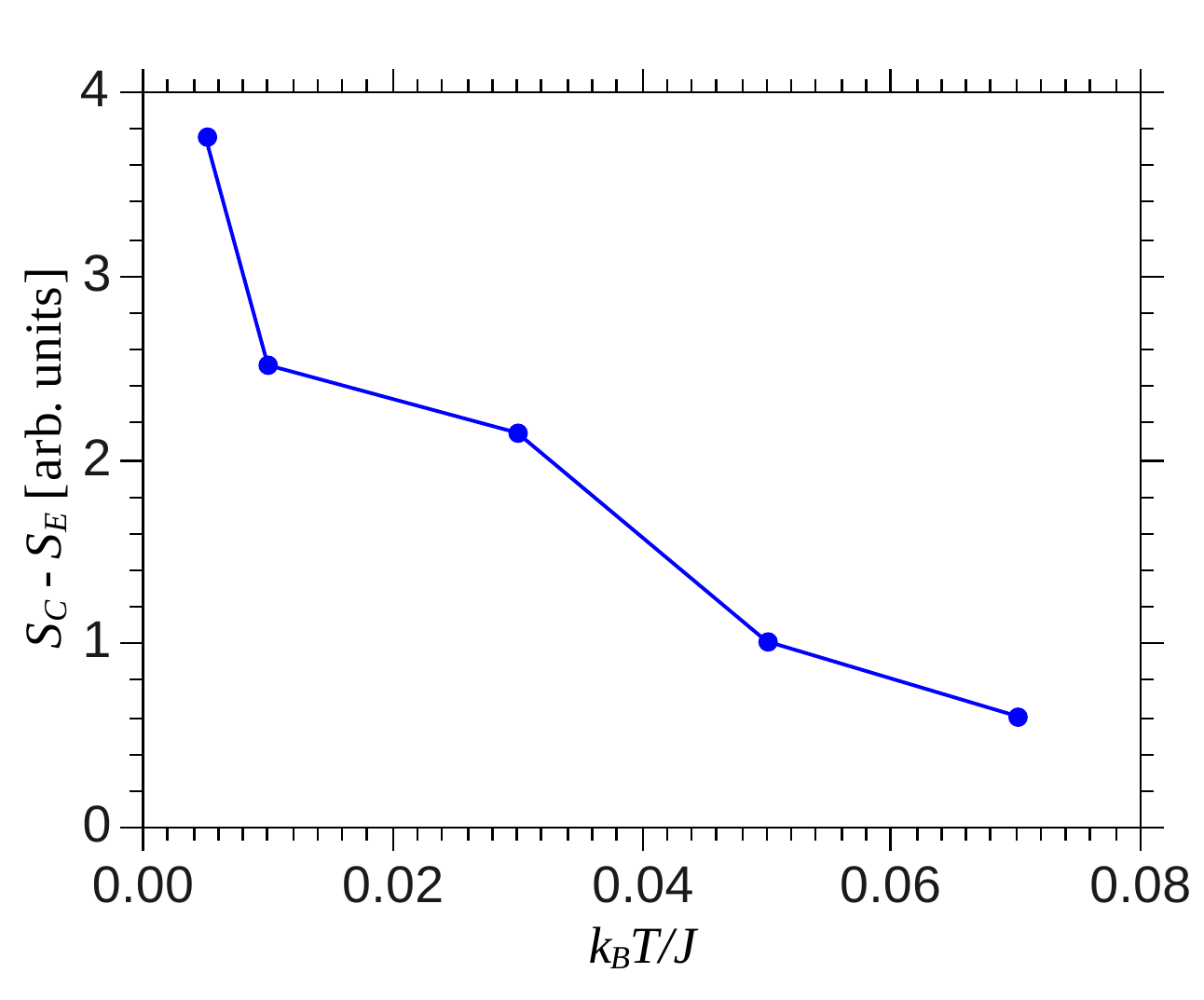}
\caption{\label{SvT_1}(Color online) Temperature dependence of the Bragg spectra in the center of the trap.
More precisely, this figure shows $\sum_{\bm k}'\bar{S}({\bm k})-\bar{S}(0,\pi)$ where the sum is calculated over
the square marked by white dotted line in Fig. \ref{r12_9}
}
\end{figure}

It can be seen in Fig. \ref{r12_9}A, that for $R=12.9a$ and $U=J$ a checkerboard phase occupies the central part 
of the trap. It leads to a highly pronounced peak at ${\bm k}=(\pi,\pi)$. Other phases give
smaller peaks at $(\pi,\pi/2)$, $(\pi,0)$, $(2\pi/3,2\pi/3)$ (and points obtained by symmetry operations). 

For $R=18.5a$ and $U=5J$ there are no heavy atoms in the center of the trap: they are distributed in a ring
of some width with two different phases that have axial stripes on its innermost part. Since the stripes are 
invariant with respect to translations along the axis (neglecting the finite size of the system), all peaks are 
located at $k_x=0$ or $k_y=0$, namely at ${\bm k}=(0,\pi),\ (\pi,0),\ (0,\pm 2\pi/3)$ and $(\pm 2\pi/3,0)$. 
The outer part of the ring is densely filled with heavy atoms and its Fourier transform contributes to 
the peak at ${\bm k}=(0,0)$.

\subsubsection{Finite temperatures: Monte Carlo}

\begin{figure*}[htb]
\includegraphics[width=\textwidth]{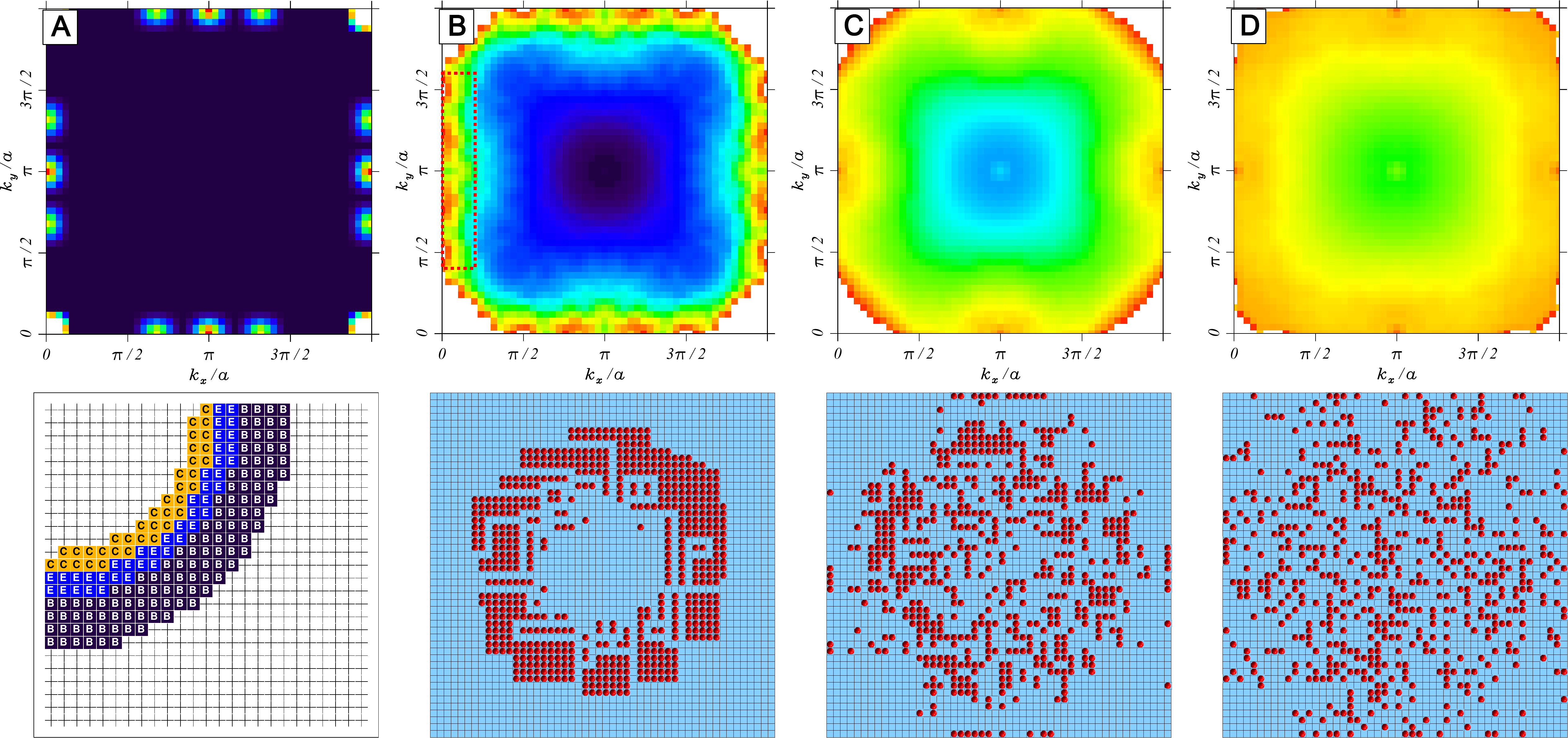}
\caption{\label{r18_5}(Color online) The same as in Fig. \ref{r12_9}, but for $U = 5J$. Panels B, C, and D
correspond to temperature $k_{\rm B}T = 0.01J,\ 0.05J,$ and $0.15J$, respectively. }
\end{figure*}

In order to investigate how temperature affects the patterns formed by the heavy atoms, Monte Carlo simulations
have been carried out. For a sufficiently long MC run, a number of independent heavy atom configurations are generated. The length of a single MC run depends on the temperature and the model parameters (interaction, shape
of the harmonic trap), but usually it is on the order of $10^6$ MC steps from which about $10^3-10^4$ 
independent configurations have been selected. For each configuration the Fourier transform has been calculated. 
Figs. \ref{r12_9}BCD and \ref{r18_5}BCD present results averaged over all configurations generated for a given 
set of model parameters. We need to comment about the averaging procedure. In an experiment, the Bragg spectra 
of a single configuration can be observed. However, in many cases the configuration (and the corresponding
spectra) changes significantly between successive ``snapshots''. In order to make our results independent of any
particular distribution of the atoms, we decided to present results that can characterize the system at a given 
temperature and model parameters. Fourier transforms of very similar configurations often have similar shape, but 
may have different sign, depending on the details, {\it e. g.}, on the phases of the order parameter for a checkerboard 
phase. Therefore, in Figs. \ref{r12_9}BCD and \ref{r18_5}BCD, we present averaged absolute values of the spectra, 
given by
\begin{equation}
\bar{S}({\bm k})=\frac{1}{MN}\sum_{m=1}^M\left|\sum_{i=1}^N w_{i,m} e^{i{\bm k}\cdot{\bm R}_i}\right|,
\end{equation}
where $N$ and $M$ are the number lattice sites and the number of MC ``snapshots'', respectively and $w_{i,m}$
is equal to 0 or 1, depending on whether site $i$ is occupied or empty in the $m$-th ``snapshot''.
Additionally, the resulting Bragg spectra are self averaged using rotational and reflection symmetries of the 
lattice and the trap. This means that the presented spectra are calculated as averages of 
$S(k_x,k_y)$, $S(k_x,-k_y)$, $S(-k_x,k_y)$, $S(-k_x,-k_y)$, $S(k_y,k_x)$, $S(k_y,-k_x)$, $S(-k_y,k_x),$ 
and $S(-k_y,-k_x)$. Since we are not interested in the ${\bm k}=(0,0)$ peak (and equivalent peaks in 
the remaining corners), the false-color scales in Figs. \ref{r12_9} and \ref{r18_5} are chosen in such
a way that (independent of the temperature) only the peaks resulting from ordering are correctly represented.
But of course, the scale is kept the same in panels B, C, and D.

\begin{figure}[htb]
\includegraphics[width=0.4\textwidth]{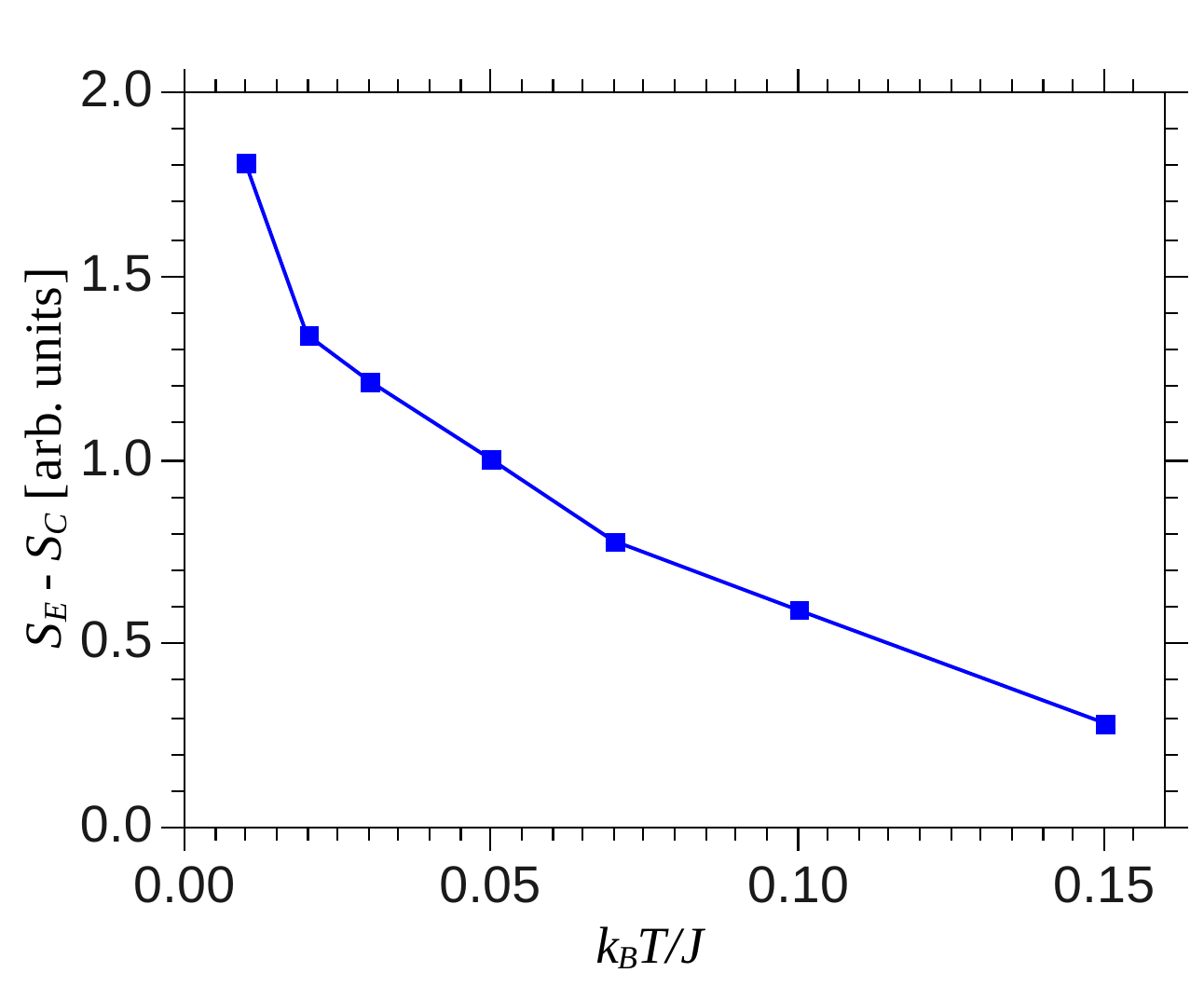}
\caption{\label{SvT_5}(Color online) Temperature dependence of the Bragg spectra at the edge of the trap:
this figure shows $\sum_{\bm k}'\bar{S}({\bm k})-\bar{S}(\pi,\pi)$ where the sum is calculated over
the rectangle marked by red dotted line in Fig. \ref{r18_5}
}
\end{figure}

In the case presented in Fig. \ref{r12_9}A (LDA, $U=J, R=12.9a$), most of the heavy atoms form a checkerboard 
configuration leading to a strong peak at ${\bm k}=(\pi,\pi)$. In the peripheral areas, the concentration
of heavy atoms increases and they form more dense patterns, which in turn, lead to less pronounced peaks.
One can notice that the central peak in Fig. \ref{r12_9}B is fourfold split. This results from an imperfection
of the checkerboard ordering of the heavy atoms. This can be explained from an analysis of the LDA: the effective
chemical potential varies continuously when the distance from the center of the trap increases. Hence one expects that the
density of atoms (both light and heavy) should also vary in a continuous way. However, in some regions, 
the checkerboard ordering minimizes the energy, hence in that ordered region the density (at least of the heavy 
atoms) is constant. A competition between these two tendencies leads to flaws in the center of the trap. 
Due to topological reasons, such a crack has to run across the entire ordered area. Then the question is 
why the fourfold split is not visible in the LDA results? This is connected with the maximum size of the 
unit cell taken into account in the LDA calculations. The $4$ site unit cell is too small to describe
a large checkerboard area with a single line defect, but by using a larger unit cell, one should see such
a configuration.

As can be expected, the ordering is reduced when the temperature increases and so are the corresponding features 
in the Bragg spectra. Figs. \ref{r12_9}C and \ref{r12_9}D illustrate how the multi-peak structure is 
smoothed out by destroying the real-space ordering. In order to describe this process in a more 
quantitative way, we calculated how the spectral weight of the central peak(s) decreases with increasing
temperature. Figure \ref{SvT_1} shows the spectral weight in a region 
$(0.75\pi\le k_x\le 1.25\pi)\times(0.75\le k_y\le 1.25\pi)$ (marked by the white square in Fig. \ref{r12_9}A)
as a function of temperature. The plot is shifted in such a way that the weight at ${\bm k}=(\pi,0)$ is 
equal to zero. The shift is necessary since with increasing temperature less and less weight is attributed 
to the peak at ${\bm k}=(\pi,0)$, which increases the reference level more than should result from the reduction 
of the central peak (this effect is especially visible in Fig. \ref{r18_5}, where the color around $(\pi,\pi)$ 
becomes very bright at high temperature).

\section{Summary}

In this paper, we have examined two simple experimental probes that can reveal information about ordered density wave phases in mixtures of light and heavy fermionic atoms (or equivalently light fermionic and heavy ``hard core'' bosonic atoms) on an optical lattice.  Namely, we examined the momentum distribution function, which comes from a time-of-flight expansion experiment and the Bragg scattering signal that would come from scattered optical light that scatters off of the density wave pattern. 

The momentum distribution function does not provide significant information about various ordered density wave phases, but it does provide some intuition about phase separated states.  In particular, as the distribution flattens and broadens, one has an indication of phase separation setting in.  Furthermore, if the heavy atoms surround the lights and confine the light atoms with essentially a hard wall boundary condition, then the momentum distribution function develops a sharp dimple at low momentum which does provide a characteristic shape signalling that form of phase separation.

Bragg scattering is much more effective at showing the presence of ordered density wave phases, as new Bragg reflection ``spots'' appear at appropriate ordering vectors for the different types of order present in the sample.  The weight underneath these peaks is proportional to the strength of the ordering, and to the volume of regions which are ordered, and hence they can be used for accurate low-temperature thermometry of these systems. As $T$ is lowered, the weight in the peaks grows and can be calibrated via numerical calculations to produce an appropriate temperature of the system.  

One caveat of this work, however, is that one must cool the mixture down to a low enough temperature that the ordering appears in the system.  Typically, this requires a temperature at least as low as about 1/40th of the bandwidth, and often substantially lower.  This is an aggressively low temperature with current cooling technology, but hopefully can be reached as it becomes easier to manipulate entropy distributions within trapped atomic systems.  Finally, we also should note that direct {\it in situ} imaging via apparatus like the quantum gas microscope would provide even more convincing pictures of the ordered phases, and the fluctuations about that order, than the above proposed methods, but we are not aware of any plans to examine these kinds of mixtures with such machines in the near term.

\begin{acknowledgments}
M.M.M. acknowledges support under Grant No. NN~202~128736 from Ministry of Science and Higher Education 
(Poland). J. K. F. acknowledges support under ARO grant number W911NF0710576 with funds from the DARPA OLE program and support from the McDevitt endowment trust at Georgetown University. We acknowledge M. Rigol for a critical reading of the manuscript.
\end{acknowledgments}

\end{document}